\documentclass[prl,twocolumn,superscriptaddress]{revtex4-2}
\usepackage{epsfig}
\usepackage{graphicx}
\usepackage{palatino}
\usepackage[english]{babel}
\usepackage{hyphenat}
\usepackage{amsmath}
\usepackage{amssymb}
\usepackage{mathtools}
\usepackage{mathrsfs}
\usepackage{bbm} 
\usepackage{bm}
\usepackage{slashed}
\usepackage{epstopdf}
\usepackage{xcolor}
\usepackage{booktabs}
\usepackage{float}
\definecolor{lcolor}{rgb}{0.,0.0,0.}
\definecolor{citcolor}{rgb}{0,0.,0.5}
\usepackage[breaklinks,colorlinks,urlcolor=blue,citecolor=blue,linkcolor=blue]{hyperref}
\usepackage{multirow}
\usepackage{ltablex}
\usepackage{soul}

\newcommand{\Hcal}{\mathcal{H}}

\newcommand{\vect}[1]{\boldsymbol{#1}_{\perp}}
\newcommand{\kt}{\vect{k}}

\newcommand{\Pt}{\vect{P}}

\newcommand{\qt}{\vect{q}}

\newcommand{\at}{\vect{a}}
\newcommand{\bt}{\vect{b}}
\newcommand{\Bt}{\vect{B}}

\newcommand{\ltone}{\boldsymbol{l_{1\perp}}}
\newcommand{\lttwo}{\boldsymbol{l_{2\perp}}}
\newcommand{\ltthre}{\boldsymbol{l_{3\perp}}}

\newcommand{\xt}{\vect{x}}
\newcommand{\yt}{\vect{y}}
\newcommand{\zt}{\vect{z}}

\newcommand{\Lt}{\vect{L}}

\newcommand{\rxyt}{\boldsymbol{r}_{xy}}

\newcommand{\rzxt}{\boldsymbol{r}_{zx}}
\newcommand{\rzyt}{\boldsymbol{r}_{zy}}
\newcommand{\rzbt}{\boldsymbol{r}_{zB}}

\newcommand{\rzpbt}{\boldsymbol{r}_{z'B}}

\newcommand{\rzzpt}{\boldsymbol{r}_{zz'}}

\newcommand{\rzpyt}{\boldsymbol{r}_{z'y}}

\newcommand{\rxxtp}{\boldsymbol{r}_{xx'}}
\newcommand{\ryytp}{\boldsymbol{r}_{yy'}}

\newcommand{\rzxpt}{\boldsymbol{r}_{zx'}}

\newcommand{\rzpxt}{\boldsymbol{r}_{z'x}}

\newcommand{\rxpyt}{\boldsymbol{r}_{x'y}}

\newcommand{\der}{\mathrm{d}}

\newcommand{\Tr}{\mathrm{Tr}}

\newcommand{\beq}{\begin{equation}}
\newcommand{\eeq}{\end{equation}}
\newcommand{\bal}{\begin{align}}
\newcommand{\eal}{\end{align}}

\begin{document}

\title{Small-$x$ factorisation in the target fragmentation region}
\author{Paul Caucal}
\email{caucal@subatech.in2p3.fr}
 \affiliation{SUBATECH UMR 6457 (IMT Atlantique, Université de Nantes, IN2P3/CNRS), 4 rue Alfred Kastler, 44307 Nantes, France}
 \author{Farid Salazar}
 \email{farid.salazar@temple.edu}

\affiliation{Department of Physics, Temple University, Philadelphia, PA 19122 - 1801, USA}

 \affiliation{RIKEN-BNL Research Center, Brookhaven National Laboratory, Upton, New York 11973, USA}

 \affiliation{Physics Department, Brookhaven National Laboratory, Upton, New York 11973, USA}

 \affiliation{Institute for Nuclear Theory, University of Washington, Seattle WA 98195-1550, USA}

\begin{abstract}
We consider the differential cross-section for single-inclusive jet production with transverse momentum $P_\perp$ in Deep Inelastic Scattering (DIS) at small Bjorken $x_{\rm Bj}$, mediated by a virtual photon with virtuality $Q^2$. Unlike most studies at small $x$, which focus on particle production in the current fragmentation region, we investigate the kinematic regime where the jet is produced in the target fragmentation hemisphere of the Breit frame, and with $P_\perp \ll Q$. For a longitudinally polarised photon, we demonstrate that this cross-section is not power suppressed in $P_\perp/Q$ and we derive a factorised expression in terms of extended quark and gluon jet fracture functions. Our formula, valid at next-to-leading order in $\alpha_s$ at small $x$, is akin to the Altarelli-Martinelli identity for the longitudinal DIS structure function. Numerical estimates show that the extended quark jet fracture function is the most sensitive to saturation effects in large nuclei.
\end{abstract}

\maketitle

Transverse momentum dependent (TMD) distribution functions~\cite{Boussarie:2023izj}, which characterise the distribution of quarks and gluons inside nuclei in terms of their longitudinal and transverse momenta with respect to the nucleus direction of motion, provide fundamental insight on hadron substructure bounded in nuclei~\cite{Arleo:2025oos}. As such, their measurement is one of the main goals of the Electron-Ion Collider (EIC) scientific program~\cite{Accardi:2012qut,AbdulKhalek:2021gbh,Achenbach:2023pba}. 
A key process to measure the quark TMDs is semi-inclusive hadron production in Deep Inelastic Scattering (DIS) in the regime where the transverse momentum squared $P_\perp^2$ of the hadron (measured in the Breit frame~\cite{Feynman:356451,Streng:1979pv,Rinehimer:2009yv}) is much smaller than the virtuality $Q^2$ of the photon probing the target wavefunction~\cite{Tangerman:1994eh,Mulders:1995dh,Boer:1997nt}.

Most studies of semi-inclusive DIS (SIDIS) in the TMD literature have focused on hadrons measured in the current (i.e.\,virtual photon) fragmentation region of the Breit frame for which a factorisation theorem has been established at leading power (LP) in $P_\perp/Q$~\cite{Collins:1992kk,Ji:2004wu,Collins:2011zzd}. On the other hand, the case where the hadron is measured in the target fragmentation region (TFR) has received less attention, although the corresponding cross-section in the Bjorken limit (fixed Bjorken variable $x_{\rm Bj}\lesssim 1$, large $Q^2$) admits a factorisation theorem~\cite{Grazzini:1997ih,Collins:1997sr} formulated in terms of extended fracture functions~\cite{Trentadue:1993ka,Graudenz:1994dq,deFlorian:1997wi,Ceccopieri:2007th}. These extended fracture functions describe the distribution of partons carrying longitudinal momentum fraction $x$ in the target probed by the virtual photon under the condition that a spectator parton fragments into a hadron/jet in the TFR with a given longitudinal momentum fraction of the target $\xi$ and transverse momentum $P_\perp$. Fracture functions thus provide additional insight into the internal parton dynamics and, as such, they have recently been measured at moderate $x_{\rm Bj}$~\cite{Ceccopieri:2014rpa,CLAS:2022sqt}. When the hadron measured in the TFR is the recoiling target nucleon in diffractive events, extended fracture functions share the same operator definition~\cite{Hautmann:1999ui} and thus contain the same information as diffractive parton distribution functions (PDF)~\cite{Collins:1997sr,Berera:1994xh}; however, fracture functions are more general objects that have never been computed at small-$x$.

In this paper, we investigate for the first time jet production in SIDIS in the TFR and in the Regge limit (small $x_{\rm Bj}\ll1$, fixed~$Q^2$)~\cite{Regge:1959mz}. From a broader perspective, our main goal is to bring attention to the TFR as a key probe of saturation dynamics, as the transverse momentum of the spectator parton produced in the TFR reflects the saturation imprint of its parent gluon in a perturbatively calculable manner, positioning the TFR as a valuable region for saturation study at the EIC. This study is also motivated by the connection~\cite{Chen:2024bpj} between fracture functions and nucleon energy-energy correlators~\cite{Liu:2022wop}, a novel observable proposed to search for saturation effects at small $x$~\cite{Liu:2023aqb}. 
 
For the sake of simplicity, we consider the longitudinally polarised photon contribution to SIDIS. An analogous analysis can be carried out for the transversely polarized case. Surprisingly, unlike hadron/jet measured in the current hemisphere, which give power suppressed contributions to the longitudinal structure function~\cite{Mulders:1995dh,Bacchetta:2006tn,Vladimirov:2021hdn,Rodini:2023plb,Ebert:2021jhy}, we demonstrate here (i) that the single-inclusive jet cross-section in SIDIS mediated by a longitudinally polarised photon is LP (i.e.~not suppressed by $P_\perp/Q$ for $Q\gg P_\perp$) starting from next-to-leading order (NLO) in the strong coupling $\alpha_s$, and controlled by jets produced in the TFR, (ii) that this LP contribution factorises in terms of extended quark and gluon jet fracture functions for which we give explicit analytic expressions in the saturation regime $P_\perp\sim Q_s$ with $Q_s$ the nucleus saturation scale.

Our strategy to achieve this is simple: we shall extract the LP contribution to the single-inclusive jet cross-section in longitudinally polarised DIS based on the NLO formula \textit{valid for general jet kinematics} at small $x$. A byproduct of this calculation will be the proof that the LP term in the $P_\perp/Q$ expansion comes indeed from jets produced in the target hemisphere. The benefit of using jets is to avoid non perturbative inputs like final state fragmentation functions~\cite{Dumitru:2005gt,Caucal:2024nsb}, although the phase space for producing jets with large enough $P_\perp$ while being in the regime $P_\perp\ll Q,x_{\rm Bj}\ll 1$ is quite narrow at the EIC (the situation will be more favourable at future high energy $e^-$-$p/A$ colliders~\cite{LHeCStudyGroup:2012zhm,FCC:2018byv,LHeC:2020van,Bruning:2260408}).A more systematic hadron vs.~jet analysis should be conducted using suitable jet definitions~\cite{Catani:1992zp,Webber:1993bm,Cacciari:2011ma,Caucal:2024vbv,Arratia:2020ssx} and handling non-perturbative effects.

We now introduce the single-inclusive jet cross-section, in the one virtual photon ($\gamma_{T/L}^*$) exchange approximation~\cite{Bacchetta:2006tn},
\begin{align}
    \frac{\der\sigma^{e^-+A\to {e^-}'+\textrm{jet}+X}}{\der x_{\rm Bj}\der Q^2\der^2\Pt}&=\sigma_{0}^{\rm DIS}\left[F_{UU,T}+\varepsilon F_{UU,L}\right]\,,\label{eq:SIDIS-jet-xs}
\end{align}
where $\sigma_{0}^{\rm DIS}=4\pi\alpha_{\rm em}^2(1-y+y^2/2)/(x_{\rm Bj}Q^4)$ is the Born level hard factor and $\varepsilon=(1-y)/(1-y+y^2/2)$ is the ratio between the longitudinal and transverse photon fluxes. $Q^2$, $x_{\rm Bj}$ and $y=Q^2/(sx_{\rm Bj})$ respectively denote the virtuality, the Bjorken variable, and the inelasticity for a fixed $e^-+A$ center-of-mass energy per nucleon $\sqrt{s}$. $\alpha_{\rm em}$ is the fine structure constant. $F_{UU,T}$ and $F_{UU,L}$ are respectively the transverse and longitudinal contributions to the single-inclusive jet cross-section. 
In the limit $P_\perp \ll Q$, $F_{UU,T}$ is leading-twist at tree-level and the cross-section for transversely polarised $\gamma_T^*$ admits TMD factorisation; the NLO calculation of $F_{UU,T}$ for jets at moderate $x_{\rm Bj}$ can be found in~\cite{Caucal:2024vbv}.

As a proof of principle, we shall compute $F_{UU,L}(x_{\rm Bj},Q^2,\Pt)$ in Eq.\,\eqref{eq:SIDIS-jet-xs} in the small $x_{\rm Bj}$ limit. This is sufficient to prove that $F_{UU,L}$ is LP. The calculation at small $x_{\rm Bj}$ is performed in the Colour Glass Condensate (CGC) effective field theory~\cite{Iancu:2002xk,Iancu:2003xm,Gelis:2010nm,Kovchegov:2012mbw,Morreale:2021pnn} and in the dipole frame~\cite{Kopeliovich:1981pz, Bertsch:1981py, Mueller:1989st,Nikolaev:1990ja} where the virtual photon has 4-momentum $q^\mu=(q^+,-Q^2/(2q^+),\boldsymbol{0}_\perp)$ and a nucleon from the nucleus $A$ has 4-momentum $P^\mu=(0,P^-,\boldsymbol{0}_\perp)$ in light-cone coordinates~\footnote{We neglect target mass corrections.}. Another advantage of the small $x$ limit is that the CGC provides an input for the non-perturbative part for the TMD PDF~\cite{McLerran:1998nk,Venugopalan:1999wu,Mueller:1999wm,Xiao:2017yya,Marquet:2009ca,Dominguez:2011wm,Petreska:2018cbf}. 
This also applies to extended jet fracture functions, so our calculation provides their CGC expressions at small $x$ for the first time. 
We note that Eq.\,\eqref{eq:SIDIS-jet-xs} is not differential with the jet's rapidity; including this dependence, together with the hadron production case, will be addressed in a companion paper.

\begin{figure}
    \centering
    \includegraphics[width=0.99\linewidth]{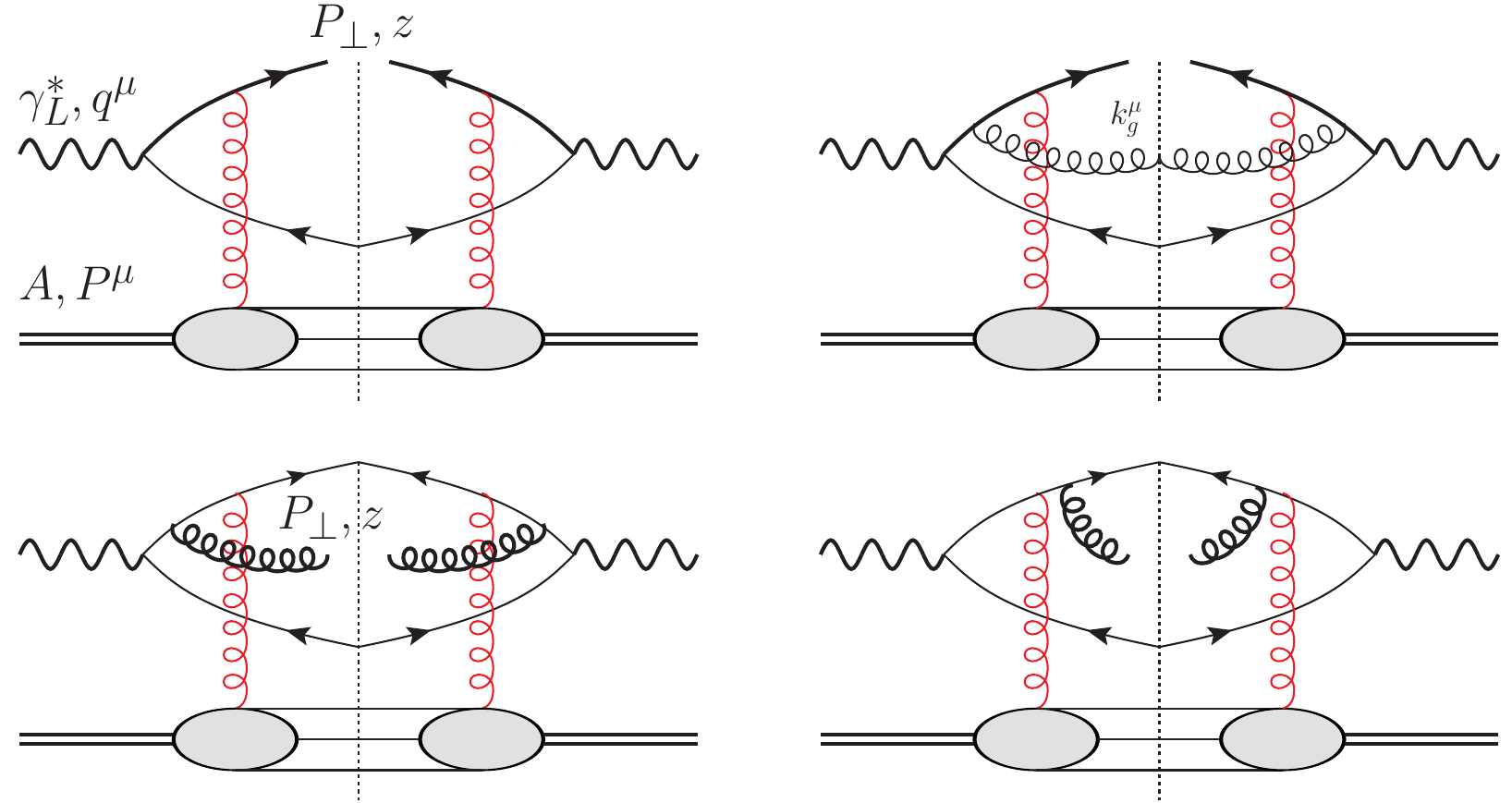}
    \caption{Feynman graphs for the $\gamma^*_L+A\to \textrm{jet}+X$ sub-process at small $x_{\rm Bj}$. Top: LO (left) and LP NLO diagram (right) for a quark jet measurement. Bottom: examples of LP NLO diagrams for gluon jet production ($q\leftrightarrow\bar q$ exchanged and interference graphs are not displayed but contribute at LP).}
    \label{fig:Feynman-graphs}
\end{figure}

The LO cross-section is given by the top-left diagram in Fig.\,\ref{fig:Feynman-graphs}, where the measured quark with four-momentum $k^\mu$ and fractional electric charge $e_q$ has a longitudinal momentum fraction $z=k\cdot P/(P\cdot q)=k^+/q^+$ and transverse momentum $\Pt=\kt$ w.r.t.~the photon direction. In this diagram, multiple scattering of the quark-antiquark colour dipole with transverse size $\rxyt$ off the shockwave --- pictured by a red gluon in Fig.\,\ref{fig:Feynman-graphs} and representing the dense gluon field of the target --- is encoded into the dipole $S$-matrix, generically defined in the representation $R=F,A$ of $SU(3)$ ($F=$ fundamental, $A=$ adjoint) as 
\begin{align}
    \mathcal{D}_R(x,\qt)\equiv\int_{\rxyt}\frac{e^{-i\qt\cdot\rxyt}}{(2\pi)^2d_R}\left\langle \Tr\left[V_{R,\xt}V^\dagger_{R,\yt}\right]\right\rangle_x\,,\label{eq:dipole-def}
\end{align}
with $V_{R,\xt}$ a light-like Wilson line in representation $R$ with transverse coordinate $\xt=\Bt+\rxyt/2$ and $d_R$ the dimension of $R$. $V_{F,\xt}$ describes the colour precession of the quark while eikonally interacting with the target field at $\xt$ (resp.~$\yt=\Bt-\rxyt/2$ for the antiquark). We use the shorthand notation $\int_{\rxyt}=\int\der^2\rxyt$. The dependence upon the impact parameter $\Bt$ of the dipole in Eq.\,\eqref{eq:dipole-def} is omitted. The $x$ dependence of the CGC average $\langle ...\rangle_x$ over gluon field configurations is determined by the BK-JIMWLK evolution equation~\cite{Balitsky:1995ub,Kovchegov:1999yj,JalilianMarian:1997jx,JalilianMarian:1997gr,Kovner:2000pt,Weigert:2000gi,Iancu:2000hn,Iancu:2001ad,Ferreiro:2001qy} or by its linear version corresponding to the BFKL equation~\cite{Kuraev:1977fs,Balitsky:1978ic} in the limit $q_\perp\gg Q_s$, with $Q_s$ the nucleus saturation scale. Ultimately, $x$ is fixed by four-momentum conservation at the scale $x=x_{\rm Bj}$. Gathering the dipole $S$-matrix and the hard factor $\Hcal_{\rm LO}$ for producing the $q\bar q$ pair from the virtual photon, we get \cite{Mueller:1999wm,Iancu:2020jch}
\begin{align}
    F^{\rm LO}_{UU,L} &=\frac{N_c}{2\pi^4} \sum_{i=q}e_i^2 \int_{\Bt,\qt} \!\!\!\!\!\! \mathcal{D}_F(x,\qt)\nonumber\\
    &\times \int_0^{1}\der z\left|\frac{\bar Q^2}{(\Pt-\qt)^2+\bar Q^2}-\frac{\bar Q^2}{P_\perp^2+\bar Q^2}\right|^2\,,\label{eq:FuuL-alltwist}
\end{align}
where the sum runs over light quark flavours and $\bar Q^2=z(1-z)Q^2$.
At this stage, Eq.\,\eqref{eq:FuuL-alltwist} contains all kinematic $\mathcal{O}(P_\perp/Q)$ and saturation $\mathcal{O}(Q_s/Q)$ power corrections at small $x$~\cite{Altinoluk:2019wyu}. Performing the integral over $z$ and expanding the result  in the limit $Q\gg P_\perp$ yields $F_{UU,L}^{\rm LO}=\mathcal{O}(P_\perp^2/Q^2)$ --- the explicit expression for the subleading power contribution from $F_{UU,L}^{\rm LO}$ is shown in the Supplemental Material (SM)~\footnote{See Supplemental Material, which includes Ref.\,\cite{Roy:2019hwr,Aschenauer:2017oxs,Kowalski:2003hm,Rezaeian:2012ji,Armesto:2022mxy,Kaptari:2019lfj,Altinoluk:2020oyd,Altinoluk:2021lvu,Altinoluk:2022jkk,Boussarie:2020fpb,Boussarie:2021wkn,Boussarie:2023xun,Fu:2023jqv,Fu:2024sba}, for more details on the derivation of Eq.~\eqref{eq:FUUL-LP-final}.} (see also~\cite{Badelek:1996ap,Badelek:2022cgr,Chen:2023wsi}).
In this limit, the integral over $z$ in Eq.\,\eqref{eq:FuuL-alltwist} is dominated by the two endpoints $z=0$ and $z=1$, that is by the so-called aligned jet configurations~\cite{Mueller:1999wm,Marquet:2009ca,Iancu:2020jch,Caucal:2024vbv} where the produced $q\bar q$ pair is highly asymmetric in longitudinal space with either $z\sim P_\perp^2/Q^2$ or $1-z\sim P_\perp^2/Q^2$. The overall $z^2(1-z)^2$ factor, originating from the $\bar Q^4$ dependence in Eq.\,\eqref{eq:FuuL-alltwist}, renders the LO cross-section power suppressed.

Physically, a longitudinally polarised photon cannot directly couple to a (sea) quark at LO from the target because of helicity conservation~\cite{Callan:1969uq,Ellis:1996mzs}. The same argument applies for the longitudinal structure function $F_L(x,Q^2)$ which also vanishes at LO in the naive parton model by virtue of the Callan-Gross relation~\cite{Callan:1969uq}. That said, we note in anticipation of a future discussion that the \textit{integral} over $\Pt$ of $F_{UU,L}^{\rm LO}$ in Eq.\,\eqref{eq:FuuL-alltwist} nevertheless contributes to $F_L$ at LP: one gets $F_L^{\rm LO}=\alpha_s(\Sigma_i e_i^2)/(3\pi)xg(x,Q^2)$~\cite{Bartels:2000hv,Golec-Biernat:2009mod,Boroun:2010zza,Boroun:2020fxg} (cf.~SM), where the gluon PDF is defined, for $x\ll 1$, as~\cite{Mueller:1999wm,Baier:1996sk,Dominguez:2010xd,Dominguez:2011wm}
\begin{align}
    xg(x,Q^2)\equiv\int^{Q^2}\der^2\qt \int_{\Bt}\frac{N_c}{2\pi^2\alpha_s}q_\perp^2 \mathcal{D}_F(x,\qt)\,.\label{eq:gluon-pdf-def}
\end{align}

We now consider the NLO correction to Eq.\,\eqref{eq:FuuL-alltwist}. The complete NLO correction at small $x_{\rm Bj}$ to the SIDIS jet cross-section for longitudinally polarised virtual photon has been computed in \cite{Caucal:2021ent,Caucal:2024cdq} (see also \cite{Ayala:2016lhd,Ayala:2017rmh,Bergabo:2022zhe}). We shall focus on the top right real diagram in Fig.\,\ref{fig:Feynman-graphs}, labelled $\rm R1\times R1^*$, 
where the quark-jet $P_\perp$ is measured (the case where the antiquark jet is tagged is identical). This diagram contributes to $F_{UU,L}$ as (cf.~Eq.\,(3.33) in \cite{Caucal:2024cdq}),
\begin{align}
     &F^{\rm R1\times R1^*}_{UU,L}=\frac{\alpha_sN_c}{16\pi^8} \sum_{i=q}e_i^2 \int_{\xt,\xt',\yt,\zt} \!\!\!\!\!\!\!\!\!\!\!\! e^{-i\Pt\cdot\rxxtp}\int_0^1\der z\nonumber\\
     &\times\int_0^{1-z}\frac{\der z_g}{z_g} \left(1-z-z_g\right)^2  Q^4 K_0(QX_R)K_0(QX_R')\nonumber\\
     &\times\frac{\rzxt \cdot\rzxpt}{\rzxt^2 \rzxpt^2}\left[z^2+ (z+z_g)^2 \right]\Xi(\xt,\yt,\zt,\xt')\,,\label{eq:FULL-R1R1-alltwist}
\end{align}
with $\boldsymbol{r}_{ab}\equiv\at-\bt$. Here $z_g=k_g\cdot P/(P\cdot q)$ is the longitudinal momentum fraction of the gluon with 4-momentum $k_g^\mu$ and $\zt$ its transverse coordinate when it eikonally crosses the shockwave. The integral over $z_g$ is divergent as $z_g\to 0$, yet as we shall see, the high virtuality limit of Eq.\,\eqref{eq:FULL-R1R1-alltwist} is insensitive to this rapidity divergence~\footnote{For general kinematics, this divergence is cancelled by the high energy evolution of the dipole $S$-matrix~\cite{Caucal:2024cdq}}.
Further, $K_0$ is the zeroth order modified Bessel function entering into the $q\bar q g$ light-cone wave-function with the effective $q\bar qg$ size given by $X_R^2=z(1-z-z_g)\rxyt^2+zz_g\rzxt^2+(1-z-z_g)z_g\rzyt^2$ and likewise for $X_R'$ with the replacement $\rxyt\to\rxpyt$ and $\rzxt\to\rzxpt$. Last, the CGC correlator $\Xi$ associated with the eikonal interaction of the $q\bar q g$ system with the target gluon field reads
\begin{align}
    \Xi=&\frac{N_c}{2}\left\langle D_{xx'}-D_{xz}D_{zy}-D_{yz}D_{zx'}+1\right\rangle_x\nonumber\\
    &-\frac{1}{2N_c}\left\langle D_{xx'}-D_{xy}-D_{yx'}+1\right\rangle_x\,,\label{eq:XINLO4}
\end{align}
with the shorthand notation $D_{xy}=\Tr(V_{F,\xt}V^\dagger_{F,\yt})/N_c$. 

We would like to demonstrate that Eq.\,\eqref{eq:FULL-R1R1-alltwist} has a leading twist component in the limit $Q^2\gg P_\perp^2$. As previously mentioned, this limit is dominated by aligned jet configurations, with end points $z \sim 0$ and $1-z-z_g \sim 0$ corresponding to soft quark and soft antiquark respectively. Yet for this diagram and photon polarization, the end-point $1-z-z_g \sim 0$  is power suppressed due the factor $(1-z-z_g)^2$ in Eq.\,\eqref{eq:FULL-R1R1-alltwist}. Let us then focus on the endpoint $z  \sim 0 $ with the scaling $z\sim P_\perp^2/Q^2\ll 1$. In this regime, one can set $z=0$ in the third line of Eq.\,\eqref{eq:FULL-R1R1-alltwist}. The additional power of $z_g^2$ cancels the light-cone divergence in $1/z_g$ of the $z_g$ integral such that one can safely integrate over $z_g$ between 0 and 1, with typically $z_g\sim 1/2$.

Furthermore, the hierarchy of scales $Q\gg P_\perp$ must be reflected in a hierarchy of transverse distances among the $q\bar q g$ system while interacting with the shockwave. From the phase in Eq.\,\eqref{eq:FULL-R1R1-alltwist}, we have $r_{xx'}\sim 1/P_\perp$ (and likewise for $r_{xy}$ and $r_{x'y}$ since $\rxxtp=\rxyt-\rxpyt$) such that the $q\bar q$ pair is widely separated in the amplitude and complex conjugate amplitude. Moreover, the $\bar q g$ pair must have a small size as the condition $X_R\lesssim 1/Q$ from the $K_0$ Bessel function implies $r_{zy}\sim 1/Q$. We can therefore approximate $X_R^2 \approx z\rxyt^2+z_g(1-z_g)\rzyt^2$ (and likewise for $X_R'^2$) where it is now easy to verify that both terms in $X_R$ are of the same order. Similarly, the Weizsäcker–Williams gluon emission kernel $\rzxt^i/\rzxt^2$ in Eq.\,\eqref{eq:FULL-R1R1-alltwist} is approximated by $-\rxyt^i/\rxyt^2$. Finally, using $\zt\sim \yt$, the CGC correlator Eq.\,\eqref{eq:XINLO4} simplifies into a sum of dipoles, with an overall $C_F$ Casimir factor:
\begin{align}
    &\Xi \approx C_F\left\langle D_{xx'}-D_{xy}-D_{yx'}+1\right\rangle_x\,,\\ & \!\!\!\! =C_F\int_{\qt}\!\!\! \left(e^{i\qt\cdot\rxyt}-1\right)\left(e^{-i\qt\cdot\rxpyt}-1\right)\mathcal{D}_F(x,\qt)\,.
\end{align}
Under these approximations, one can analytically perform the transverse coordinate integrals in Eq.\,\eqref{eq:FULL-R1R1-alltwist}, and extract the $Q\gg P_\perp$ limit of the final $z$ integration (cf.~SM), such that
\begin{align}
    &F^{\textrm{NLO,LP-}q}_{UU,L}=\frac{\alpha_s C_F}{2\pi} \sum_{i=q}e_i^2 \times \frac{N_c}{4\pi^4}\int_{\Bt,\qt}\mathcal{D}_F(x,\qt)\nonumber\\
    &\times\left[1-\frac{\Pt\cdot(\Pt-\qt)}{P_\perp^2-(\Pt-\qt)^2}\ln\left(\frac{P_\perp^2}{(\Pt-\qt)^2}\right)\right]\,,\label{eq:FUUL-NLO-final}
\end{align}
which is the complete NLO result at LP and small $x$ for the quark-jet contribution to $F_{UU,L}$ (using similar arguments, one shows that other diagrams not displayed in Fig.\,\ref{fig:Feynman-graphs} are not LP).
This expression neglects power corrections in $P_\perp/Q$, but it resums to all orders saturation power corrections in $Q_s/P_\perp$. It is then applicable in the saturation regime $P_\perp\sim Q_s$ so long as $P_\perp, Q_s \ll Q$. 

\begin{figure}
    \centering
    \includegraphics[width=0.95\linewidth]{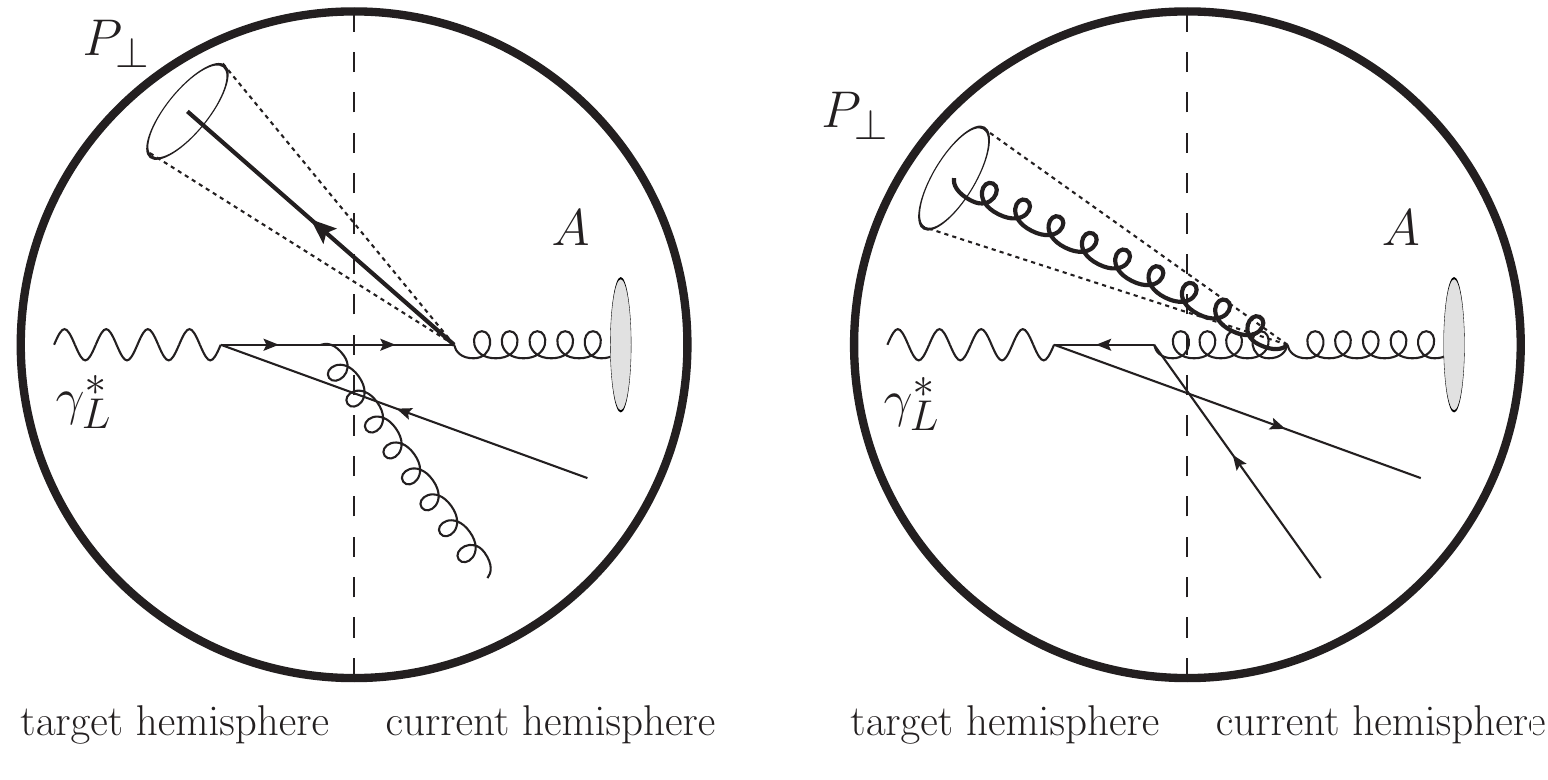}
    \caption{Breit frame views of the LP jet produced in $\gamma_L^*$-$A$ collision at small $x$ in the quark (left) and gluon (right) channels.}
    \label{fig:cartoon}
\end{figure}

Most importantly, the NLO correction to $F_{UU,L}$ is leading twist; unlike Eq.\,\eqref{eq:FuuL-alltwist}, it is not suppressed by $P_\perp/Q$. The mathematical derivation gives physical insight on why it is so: the leading twist component comes from the regime $z\sim P_\perp^2/Q^2\ll 1$, meaning that in the Breit frame where the jet rapidity reads $\eta=\ln(zQ/P_\perp)\sim \ln(P_\perp/Q)<0$, the jet is produced in the target fragmentation hemisphere (cf~Fig~\ref{fig:cartoon}-left) and not in the current (or $\gamma^*$) fragmentation one where the leading hadron is measured in hadronic SIDIS and where the corresponding TMD factorisation theorem holds~\cite{Collins:2011zzd,Boglione:2016bph,Vladimirov:2021hdn,Ebert:2021jhy}.

In Eq.\,\eqref{eq:FUUL-NLO-final}, the expression after the $\times$ symbol is interpreted as the LO quark extended jet fracture function:
\begin{align}
   &x\mathcal{F}_q(x,\Pt)=\frac{N_c}{4\pi^4}\int_{\Bt,\qt}\mathcal{D}_F(x,\qt)\nonumber\\
   &\times \left[1-\frac{\Pt\cdot(\Pt-\qt)}{P_\perp^2-(\Pt-\qt)^2}\ln\left(\frac{P_\perp^2}{(\Pt-\qt)^2}\right)\right]\label{eq:quark-TMD-FF}\,.
\end{align}
Strictly speaking, this is the $\xi$-integrated version of the extended fracture function~\cite{Grazzini:1997ih}, but the dependence on $\xi=k^-/P^-$ can easily be restored, see Eq.\,(37) in the SM. Because the jet produced in the TFR arises from perturbative parton splitting and does not come from the  intact recoiling target in a diffractive event with a rapidity gap, the $P_\perp$-dependence of the extended fracture functions differs from the $t$-dependence of diffractive PDFs computed in~\cite{Wusthoff:1997fz,Buchmuller:1998jv,Golec-Biernat:1999qor,Berera:1995fj,Hautmann:1998xn,Hautmann:1999ui,Hautmann:2000pw,Hautmann:2007cx}. Indeed, at very low values of $P_\perp \lesssim 1/R_A$ (where $R_A$ is the transverse size of the target) and when the hadron produced is the same species as the target, the diffractive SIDIS contribution is indistinguishable and must be taken into account. For values of $P_\perp \gtrsim 1/R_A$, the probability that the nucleus remains intact is negligible, and the mechanism of extended fracture functions, presented in this Letter, is the dominant contribution. Eq.\,\eqref{eq:quark-TMD-FF} turns out to be identical, at this $\alpha_s$ order, to the standard TMD sea quark distribution function probed in small $x$ SIDIS mediated by a transversely polarised photon in the current fragmentation region ~\cite{McLerran:1998nk,Venugopalan:1999wu,Mueller:1999wm,Xiao:2017yya,Marquet:2009ca,Tong:2022zwp}, including in the saturation regime $P_\perp\sim Q_s$. This is non trivial as the longitudinal photon does not directly couple to sea quarks in the target, as illustrated in Fig.\,\ref{fig:cartoon}-left.

In addition to the diagrams where the fermionic  jet is tagged, there are LP NLO contributions to the SIDIS jet cross-section where the gluon jet with transverse momentum $\Pt=\boldsymbol{k}_{g\perp}$ and $z=z_g\lesssim P_\perp^2/Q^2$ is measured~\cite{Chen:2024brp} (examples of diagrams are shown in the second row of Fig.\,\ref{fig:Feynman-graphs}). They are computed in detail in the SM following the method outlined above. Remarkably, the LP component of these diagrams now depends on the gluon-gluon dipole $\mathcal{D}_A(x,\qt)$. Indeed, in the limit $Q\gg P_\perp$ with $P_\perp$ now that of the gluon, the $q\bar q$ pair becomes the smallest dipole with transverse size $r_{xy}\sim 1/Q$ much smaller than the transverse distance $r_{zx}\sim 1/P_\perp$ between the $q\bar q$ pair and the gluon. Like in semi-inclusive diffractive DIS~\cite{Iancu:2021rup,Hatta:2022lzj}, an effective gluon-gluon dipole is built in the amplitude out of the NLO gluon and this very small $q\bar q$ dipole. The gluon-jet contribution eventually reads
\begin{align}
    F_{UU,L}^{\textrm{NLO,LP-}g}&= \frac{\alpha_s }{3\pi} \sum_{i=q}e_i^2 \times x\mathcal{F}_g(x,\Pt)\,,\label{eq:FUUL-gjet-final}
\end{align}
with the gluon extended jet fracture function $x\mathcal{F}_g$:
\begin{widetext}
\begin{align}
 x\mathcal{F}_g(x,\Pt)= \frac{N_c^2-1}{4\pi^4}\int_{\Bt,\qt}\mathcal{D}_A(x,\qt)\times&\left\{\ln\left(\frac{1}{x}\right)\frac{q_\perp^2}{P_\perp^2}-1-\frac{(\Pt-\qt)^2-q_\perp^2}{2P_\perp^2}\ln\left(\frac{(\Pt-\qt)^2}{P_\perp^2}\right)\right.\nonumber\\
+&\left.\left(1-\frac{2(\Pt\cdot(\Pt-\qt))^2}{P_\perp^2(\Pt-\qt)^2}\right)\frac{(\Pt-\qt)^2}{P_\perp^2-(\Pt-\qt)^2}\ln\left(\frac{(\Pt-\qt)^2}{P_\perp^2}\right)\right\}\,.\label{eq:gluon-TMD-FF}
\end{align}
\end{widetext}

As a cross-check of Eq.\,\eqref{eq:gluon-TMD-FF}~\cite{Chen:2021vby}, we consider the dilute limit $P_\perp\gg q_\perp\sim Q_s$ yielding, up to $Q_s^4/P_\perp^4$ corrections,
\begin{align}
    x\mathcal{F}_g(x,\Pt)&\approx \frac{\alpha_s}{2\pi^2}\frac{1}{P_\perp^2}\int_{x}^1\der\xi P_{gg}(\xi)xg(x,P_\perp^2)\,,\label{eq:Fg-dilute}
\end{align}
with $P_{gg}(\xi)=2N_c(1-\xi(1-\xi))^2/[\xi(1-\xi)_+]$ the $g\to gg$ splitting function. Thus, the gluon extended fracture function in Eq.\,\eqref{eq:gluon-TMD-FF} can be interpreted~\cite{Hauksson:2024bvv,Caucal:2024bae} as the convolution between the (adjoint) dipole gluon TMD $\mathcal{D}_A$ sourcing the parent gluon from the target wave-function and a TMD $g\to gg$ splitting function~\cite{Catani:1994sq,Ciafaloni:2005cg,Hautmann:2012sh,Hentschinski:2017ayz}, where one daughter gluon gives the measured jet while the other interacts with the virtual photon via a $q\bar q$ pair, as shown in Fig.\,\ref{fig:cartoon}-right. 

Gathering the quark and gluon jet contributions, our final result for the single-inclusive jet cross-section in longitudinally polarised DIS at LP and small $x$ is
\begin{align}
    F_{UU,L}^{\rm LP}(x,\Pt)= \ &\frac{\alpha_sC_F}{2\pi}\sum_{i=q,\bar q} e_i^2 \ x\mathcal{F}_i(x,\Pt)\nonumber\\
    &+\frac{\alpha_s}{3\pi}\sum_{i=q}e_i^2 \ x\mathcal{F}_g(x,\Pt)\,.\label{eq:FUUL-LP-final}
\end{align}
Eq.\,\eqref{eq:FUUL-LP-final}, together with Eqs.\,\eqref{eq:quark-TMD-FF}-\eqref{eq:gluon-TMD-FF}, are the main results of this paper. Eq.\,\eqref{eq:FUUL-LP-final} is analogous to the Altarelli-Martinelli identity~\cite{Altarelli:1978tq,Gluck:1978ky} for $F_L(x,Q^2)$ at \textit{one loop} in collinear factorisation, whose small $x$ limit is~\cite{Cooper-Sarkar:1987cnv,Boroun:2012bje}
\begin{align}
    F_L\underset{x\ll1}{\approx}\frac{\alpha_sC_F}{2\pi}\sum_{i=q,\bar q} e_i^2 xq_i(x,Q^2)+\frac{\alpha_s}{3\pi}\sum_{i=q}e_i^2 x g(x,Q^2)\,.\label{eq:AM-identity-main}
\end{align}
Yet, Eq.\,\eqref{eq:FUUL-LP-final} differs from Eq.\,\eqref{eq:AM-identity-main} in two crucial aspects: (i) Eq.\,\eqref{eq:FUUL-LP-final} is valid at the TMD level ($\Pt$ unintegrated), (ii) Eq.\,\eqref{eq:FUUL-LP-final} is a two loop result in the collinear factorisation $\alpha_s$ power counting since $x\mathcal{F}_{q/g}$ in Eqs.~\,\eqref{eq:quark-TMD-FF}-\eqref{eq:gluon-TMD-FF} implicitly contains one power of $\alpha_s$ through the dipole operator. This new result would be worth confirming via a two-loop collinear calculation, following~\cite{Chen:2024brp}.

Integrating Eq.\,\eqref{eq:FUUL-LP-final} for $P_\perp\le Q$ gives the one loop correction to $F_{L}$ at small $x$ and leading twist. Albeit $\mathcal{O}(\alpha_s^2)$, the result has the same structure as Eq.\,\eqref{eq:AM-identity-main} with the quark PDF $xq_i(x,Q^2)$ at small $x$ defined as the integral over $P_\perp$ up to $Q$ of $x\mathcal{F}_q$~\cite{Ebert:2022cku,delRio:2024vvq,Caucal:2024bae,Caucal:2024vbv}. In the gluon channel, $x g(x,Q^2)$ is now defined by the $\Pt$-integral of Eq.\,\eqref{eq:gluon-TMD-FF}. This integral is controlled by the perturbative $1/P_\perp^2$ tail obtained in Eq.\,\eqref{eq:Fg-dilute} and gives what one expects: the $x\to 0$ limit of one DGLAP step of the LO gluon PDF defined in Eq.\,\eqref{eq:gluon-pdf-def}. In the end, Eq.\,\eqref{eq:FUUL-LP-final} shows that the Altarelli-Martinelli relation in NLO collinear factorisation is valid at two loops for $x\ll1$.

We finally conclude with a numerical study of these extended jet fracture functions at small $x$. 
Our goal is to illustrate the $P_\perp$ dependence of $x\mathcal{F}_{q/g}$ and to assess the importance of gluon saturation. To compute the dipole $S$-matrix in momentum space, we use the McLerran-Venugopalan model~\cite{McLerran:1993ka,McLerran:1993ni}
for a nucleus with mass number $A$, $\left\langle D_{xy}\right\rangle=\exp\left[-\frac{1}{4}A^{1/3}Q_{s,p}^2\rxyt^2\ln\left(\frac{1}{r_{xy}\Lambda}+e\right)\right]$ with parameters $Q_{s,p}^2=0.2$ GeV$^2$ for the bare proton saturation scale at $x=0.01$, $\Lambda=0.24$ GeV~\cite{Lappi:2013zma,Cheung:2024qvw}. To get the adjoint dipole in Eq.\,\eqref{eq:dipole-def}, we simply rescale $Q_{s,p}^2\to N_cQ_{s,p}^2/C_F$. These choices are meant to illustrate the general behaviour of $x\mathcal{F}_{q/g}$ with motivated values for the model parameters; we postpone phenomenological studies to future work.

\begin{figure}
    \centering
    \includegraphics[width=0.44\textwidth,page=1]{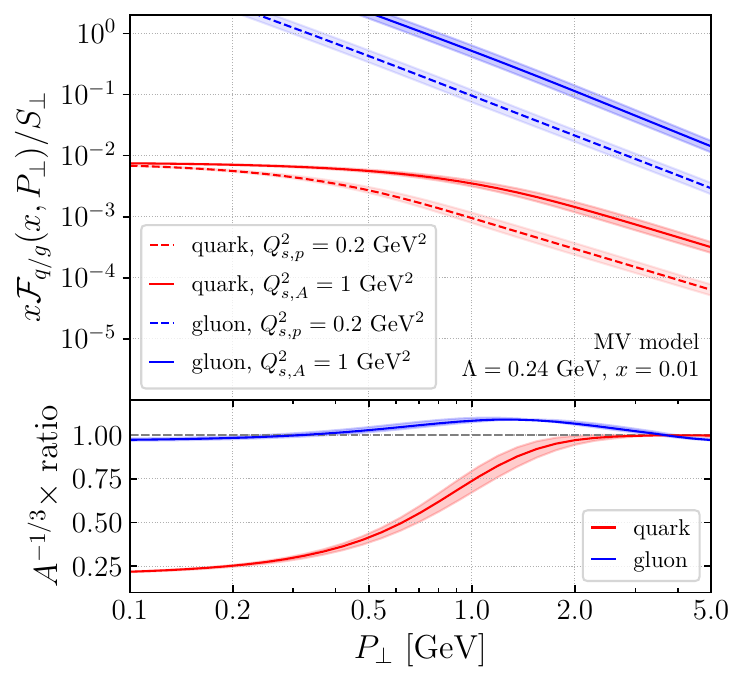}\hfill 
    \caption{Small-$x$ quark and gluon extended fracture functions at LO for a proton and a nucleus with $A^{1/3}=5$ (normalized by the transverse area $S_\perp$ of the target). The lower panel shows the nucleus/proton ratio times $A^{-1/3}$.}
    \label{fig:FULL-numeric}
\end{figure}

In Fig.\,\ref{fig:FULL-numeric}-top, the red and blue curves respectively correspond to the $P_\perp$-dependence of the quark and gluon extended fracture functions, at LO in the small $x$ limit, as given by Eqs.\,\eqref{eq:quark-TMD-FF}-\eqref{eq:gluon-TMD-FF}. They are represented for two values of the saturation scale corresponding to a proton (dashed curves) and a large nucleus with $A^{1/3}=5$ (plain curves). The bottom plot displays the ratio of the quark and gluon extended fracture functions between a large nucleus and a proton, normalised by $A^{-1/3}$, and the bands correspond to variation of $\Lambda$ by a factor of $2$.

The quark extended jet fracture function has two regimes: at large $P_\perp$, one recognises the perturbative tail $\propto A^{1/3}Q_{s,p}^2/P_\perp^2$, such that the ratio $A/p$ in the bottom plot converges towards $1$ for $P_\perp\gg Q_s$. For $P_\perp \ll Q_s$, the growth of the quark extended fracture function saturates~\cite{McLerran:1998nk,Venugopalan:1999wu,Mueller:1999wm} and reaches the universal (independent of $A$) unitary limit.

On the other hand, the gluon extended fracture function Eq.\,\eqref{eq:gluon-TMD-FF} shows very little sensitivity to saturation effect, since the $1/P_\perp^2$ perturbative tail extends down to values of $P_\perp$ smaller than $Q_s$. The small impact of saturation is also manifest in the ratio plot, which is approximatively constant except around $P_\perp\sim Q_s$.
This is a consequence of the first term in the curly bracket of Eq.\,\eqref{eq:gluon-TMD-FF}, which behaves like $1/P_\perp^2$ even for $P_\perp\ll Q_s$. This term is caused by final state gluon emissions in the dipole frame (bottom-right diagram in Fig.\,\ref{fig:Feynman-graphs}) which are insensitive to the scattering of the $q\bar q$ dipole off the shockwave. This $1/P_\perp^2$ scaling at small $P_\perp$ will be modified by quantum corrections; the extended fracture functions presented here are LO and omit quantum evolution which is essential to accurately describe their $P_\perp$ dependence. Although a rigorous proof requires a two-loop CGC calculation, we conjecture that the $x$-dependence of the fracture functions is inherited from the BK evolution of the dipole while the resolution scale $Q$-dependence should follow~\cite{Trentadue:1993ka,Chai:2019ykk} the DGLAP equation~\cite{Gribov:1972ri,Altarelli:1977zs,Dokshitzer:1977sg}. 

To summarise, we have unveiled a new contribution to small-$x$ factorisation in SIDIS coming from jets produced in the TFR. We have derived a simple analytic expression for  $F_{UU,L}$ at small $x$ and leading twist in terms of extended fracture functions. These results will open new ways to explore saturation and hadron structure phenomenology at the EIC in the TFR~\cite{Anselmino:2011ss,Anselmino:2011bb}, complementary to diffractive DIS.
As part of this program, we intend to conduct an in-depth phenomenological study of small-$x$ processes in the TFR for both jet and hadron production, including quantum evolutions of fracture functions.

\let\oldaddcontentsline\addcontentsline
\renewcommand{\addcontentsline}[3]{}
\begin{acknowledgments}\textbf{Acknowledgments.}
We would like to thank
Edmond Iancu and Feng Yuan for insightful discussions and Gholam Reza Boroun and Xuan-Bo Tong for useful comments on the manuscript. Figures~\ref{fig:Feynman-graphs} and \ref{fig:cartoon} have been created with JaxoDraw~\cite{Binosi:2003yf}. We are grateful for the support of the Saturated Glue (SURGE) Topical Theory Collaboration, funded by the U.S. Department of Energy, Office of Science, Office of Nuclear Physics. PC is funded by the Agence Nationale de la Recherche under grant ANR-25-CE31-5230 (TMD-SAT). F.S. is supported by the Laboratory Directed Research and Development of Brookhaven National Laboratory and RIKEN-
BNL Research Center. Part of this work was conducted while F.S. was supported
by the Institute for Nuclear Theory’s U.S. DOE under
Grant No. DE-FG02-00ER41132. Finally, F.S. acknowledges support from the U.S. Department of Energy, Office of Science, Office
of Nuclear Physics under the umbrella of the Quark-Gluon Tomography (QGT) Topical Collaboration with Award
DE-SC0023646.

\textbf{Data availability.} The data that support the findings of this article are openly available~\footnote{Paul Caucal and Farid Salazar, ancillary files in \href{https://arxiv.org/abs/2502.02634}{arXiv:2502.02634}.}.
\end{acknowledgments} 
\let\addcontentsline\oldaddcontentsline

\let\oldaddcontentsline\addcontentsline
\renewcommand{\addcontentsline}[3]{}
\bibliographystyle{apsrev4-1}
\bibliography{refs}
\let\addcontentsline\oldaddcontentsline

\clearpage

\appendix
\begin{widetext}

\let\oldaddcontentsline\addcontentsline
\renewcommand{\addcontentsline}[3]{}
\section{Supplemental material}
\let\addcontentsline\oldaddcontentsline

\tableofcontents
\vspace{10 mm}

\section{A. Leading twist contribution to $F_L$ from the LO structure function $F_{UU,L}^{\rm LO}$ for semi-inclusive DIS jet production}

In this Supplemental Material, we demonstrate that although $F_{UU,L}^{\rm LO}$ given by Eq.\,(3) in the main text is power suppressed by $P_\perp^2/Q^2$, the integral over $\Pt$ of Eq.\,(3) provides at leading power contribution to $F_L$ proportional to the gluon PDF. For phenomenological studies of the longitudinal structure function at small $x$, we refer the reader to \cite{Aschenauer:2017oxs,Kowalski:2003hm,Rezaeian:2012ji,Armesto:2022mxy,Boroun:2010zza,Boroun:2020fxg,Kaptari:2019lfj}.

Eq.\,(3) in the main text contains all kinematic $\mathcal{O}(P_\perp/Q)$ and saturation $\mathcal{O}(Q_s/Q)$ power corrections at small $x$. Performing the integral over $z$ and expanding the result  in the limit $Q\gg P_\perp$ yields
\begin{align}
   &F^{\rm LO, N^2LP}_{UU,L}=\frac{N_c P_\perp^2}{\pi^4Q^2} \sum_{i=q}e_i^2 \int_{\Bt,\qt} \!\!\!\!\!\! \mathcal{D}_F(x,\qt)\left[\frac{(\Pt-\qt)^2}{P_\perp^2}+1-\frac{2(\Pt-\qt)^2}{P_\perp^2-(\Pt-\qt)^2}\ln\left(\frac{P_\perp^2}{(\Pt-\qt)^2}\right)\right]\,.\label{eq:FUUL-N2LP}
\end{align}
Clearly, because of the overall prefactor $P_\perp^2/Q^2$, the LO contribution to $F_{UU,L}$ is power-suppressed.

We now consider the contribution of Eq.\,(3) in the main text to the longitudinal structure function. By definition, at this leading perturbative order where jets and partons are in one-to-one correspondence,
\begin{equation}
    F_L(x,Q^2)\equiv\int\der^2\Pt F_{UU,L}^{\rm LO}(x,Q^2,\Pt)\,.
\end{equation}
The integral over $\Pt$ can be performed analytically, such that 
\begin{align}
    F_L(x,Q^2)&=\frac{N_c}{\pi^3} \sum_{i=q}e_i^2 \int_{\Bt,\qt}\mathcal{D}_F(x,\qt)\int_0^1\der z \  \bar Q^2\left[1-\frac{2\bar Q^2}{q_\perp^2\sqrt{1+4\bar Q^2/q_\perp^2}}\ln\left(1+\frac{q_\perp^2}{2\bar Q^2}\left[1+\sqrt{1+4\bar Q^2/q_\perp^2}\right]\right)\right]\,.
\end{align}
Expanding the quantity inside the $z$ integral for $Q^2\gg q_\perp^2\sim Q_s^2$, we get
\begin{align}
    \bar Q^2\left[1-\frac{2\bar Q^2}{q_\perp^2\sqrt{1+4\bar Q^2/q_\perp^2}}\ln\left(1+\frac{q_\perp^2\left[1+\sqrt{1+4\bar Q^2/q_\perp^2}\right]}{2\bar Q^2}\right)\right]=\frac{q_\perp^2}{6}+\mathcal{O}\left(\frac{q_\perp^4}{\bar Q^2}\right)\,,
\end{align}
which is independent of $z$ at the lowest order in the $1/Q$ expansion. This means that the $z$ integral is not controlled by the end-points $z=0$ and $z=1$.
Performing the trivial integration over $z$, we get
\begin{align}
    F_L(x,Q^2)=\frac{N_c}{6\pi^3} \sum_{i=q}e_i^2 \int_{\Bt}\int^{Q^2}\der^2\qt \ q_\perp^2 \mathcal{D}_F(x,\qt) \,,
\end{align}
where the logarithmically divergent $\qt$ integral is cut by the large virtuality scale $\qt^2\le Q^2$ in agreement with our $Q^2\gg q_\perp^2$ initial assumption. Using the definition of the gluon PDF at small $x$ given by Eq.\,(4) in the main text, one finds the gluon contribution to $F_L$,
\begin{align}
    F_L^g(x,Q^2)=\frac{\alpha_s }{3\pi} \sum_{i=q}e_i^2 xg(x,Q^2)\,,\quad\textrm{ for }x\ll 1\,.\label{eq:FL-gcontrib}
\end{align}
This is in agreement with the Altarelli-Martinelli formula for $F_L$~\cite{Altarelli:1978tq,Gluck:1978ky}:
\begin{align}
   F_L(x,Q^2)&=\frac{\alpha_s}{2\pi}  x^2\int_x^1\frac{\der z}{z^3}\left[\frac{8}{3}F_2(z,Q^2)+4\left(\sum_{i=q}e_i^2\right)\left(1-\frac{x}{z}\right)zg(z,Q^2)\right]\,,\label{eq:AM-identity}
\end{align}
with $F_2$ the standard second DIS structure function.
Focusing on the second term alone, approximating $zg(z,Q^2)\approx xg(x,Q^2)$ at small $x$, and performing the $z$ integral
\begin{align}
    \int_x^1\frac{\der z}{z^3}\left(1-\frac{x}{z}\right)=\frac{1}{6x^2}\left[1+\mathcal{O}(x^2)\right]\,,
\end{align}
we find indeed Eq.\,\eqref{eq:FL-gcontrib}.

\paragraph{The sea quark contribution to $F_L$.} Formally, Eq.\,\eqref{eq:AM-identity} is valid at $\mathcal{O}(\alpha_s)$ and therefore does not contain the contribution from sea quarks. However, one can guess what the connection should be between $F_L$ and the sea quark PDF at small $x$ by using the identity $F_2(x,Q^2)= \sum_{i=q,\bar q}e_i^2 xq_{i}(x,Q^2)$ from the naive parton model (with $xq_{i}(x,Q^2)$ the valence quark PDF) and by extending it to the case of sea quarks. For sea quarks at small $x$, one can approximate $zq(z,Q^2)$ in the $z$-integral in Eq.\,\eqref{eq:AM-identity} by $xq_{\mathrm{sea},i}(x,Q^2) = xq_{i}(x,Q^2) + x\bar{q}_{i}(x,Q^2) $ since $xq_{\mathrm{sea},i}(x,Q^2)$ grows like a power law as $x$ decreases. Using 
\begin{align}
    x^2\int_x^1\frac{\der z}{z^3}=\frac{1}{2}+\mathcal{O}(x^2) \,,
\end{align}
we get
\begin{align}
    F_L^q(x,Q^2)=\frac{2\alpha_s}{3\pi} \sum_{i=q}e_i^2 xq_{\mathrm{sea}, i}(x,Q^2)=\frac{\alpha_s C_F}{2\pi} \sum_{i=q}e_i^2 xq_{\mathrm{sea}, i }(x,Q^2)\,,\quad\textrm{ for }x\ll 1\,,\label{eq:FL-qcontrib}
\end{align}
which is akin to the relation between the LP $F_{UU,L}$ and the sea quark TMD obtained in the main text. 

\section{B. Proof of Eq.\,(9) in the main text}

In this Supplemental Material, we provide the mathematical proof of Eq.\,(9) in the Letter. Our starting point is Eq.\,(5) in the Letter. After making the approximations discussed in the main text and the change of transverse coordinate variables $\xt,\yt,\xt',\zt\to \Bt=(\xt+\xt')/2,\rxyt=\xt-\yt,\rxpyt=\xt'-\yt,\rzyt=\zt-\yt$, Eq.\,(7) becomes
\begin{align}
    F^{\rm R1\times R1^*}_{UU,L}=\frac{\alpha_sN_c}{16\pi^8} \sum_{i=q} e_i^2 \int\der^2\qt\int_{\Bt,\rxyt,\rxpyt,\rzyt}&e^{-i\Pt\cdot(\rxyt-\rxpyt)}\int_0^1\der z\int_0^{1}\der z_g z_g(1-z_g)^2Q^4K_0(QX_R)K_0(QX_R')\nonumber\\
    &\times \frac{\rxyt \cdot\rxpyt}{\rxyt^2 \rxpyt^2}\left(e^{i\qt\cdot\rxyt}-1\right)\left(e^{-i\qt\cdot\rxpyt}-1\right)C_F\mathcal{D}_F(x,\qt)\,,\label{eq:FULL-R1R1-alltwis-approx}
\end{align}
with $X_R= z\rxyt+z_g(1-z_g)\rzyt$ and $X_R'= z\rxpyt+z_g(1-z_g)\rzyt$.
The first trick is to introduce the unity in Eq.\,\eqref{eq:FULL-R1R1-alltwis-approx} with
\begin{align}
    1= \int\frac{\der^2\ltthre}{(2\pi)^2}\int\der^2\rzpyt \ e^{i\ltthre \cdot (\rzyt-\rzpyt)}=\int\der^2\rzpyt \ \delta^{(2)}(\rzyt-\rzpyt)\,,
\end{align}
which allows us to replace $\rzyt$ by $\rzpyt$ in $X_R'$. Since we work under the assumption $r_{zy} \sim 1/Q$, it means that the auxiliary momentum scale $|\ltthre|$ is also of order $Q$.
Grouping together the integrals over transverse coordinates with and without prime indices, we get
\begin{align}
    F^{\rm R1\times R1^*}_{UU,L}&=\frac{\alpha_sC_FN_c}{2\pi^5} \sum_{i=q} e_i^2  Q^2\int_0^1\der z\int_0^1\der z_g \ z_g(1-z_g)^2\int_{\Bt,\qt}\mathcal{D}_F(x,\qt)\int\frac{\der^2\ltthre}{(2\pi)} \Hcal_{\rm R1\times R1^*}^{j}\Hcal_{\rm  R1\times R1^*}^{j*}\,,\label{eq:FUUL-R1R1-app1}
\end{align}
with the ``hard factor" of diagram $\rm R1\times R1^*$ defined as
\begin{align}
    \Hcal_{\rm R1\times R1^*}^j&\equiv\int\frac{\der^2\rxyt}{(2\pi)}\int\frac{\der^2\rzyt}{(2\pi)} e^{-i\Pt\cdot\rxyt}\left(e^{i\qt\cdot\rxyt}-1\right)e^{i\ltthre\cdot\rzyt}\frac{Q\rxyt^j}{\rxyt^2}K_0\left(\Delta\sqrt{\rzyt^2+\omega\rxyt^2}\right)\,,
\end{align}
where $\Delta^2=z_g(1-z_g)Q^2$ and $\omega=z/(z_g(1-z_g))$.
The hard factor $\Hcal_{\rm R1\times R1^*}$ can be computed analytically by using the inverse Fourier transform of the $K_0$ modified Bessel function (see e.g.~\cite{Roy:2019hwr,Caucal:2021ent}):
\begin{align}
    \frac{\rxyt^j}{\rxyt^2}K_0\left(\Delta\sqrt{\rzyt^2+\omega\rxyt^2}\right)&=-i\int\frac{\der^2\ltone}{(2\pi)}\int\frac{\der^2\lttwo}{(2\pi)}\frac{\lttwo^j e^{i\ltone\cdot \rzyt+i\lttwo\cdot\rxyt}}{\left(\ltone^2+\Delta^2\right)\left[\lttwo^2+\omega\left(\ltone^2+\Delta^2\right)\right]}\,,
\end{align}
such that
\begin{align}
    \Hcal_{\rm R1\times R1^*}^j&=\frac{-i(\Pt^j-\qt^j)Q}{(\ltthre^2+\Delta^2)[(\Pt-\qt)^2+\omega(\ltthre^2+\Delta^2)]}+\frac{i\Pt^j Q}{(\ltthre^2+\Delta^2)[P_\perp^2+\omega(\ltthre^2+\Delta^2)]}\,.
\end{align}
We observe that one is not allowed to make any further approximation in this hard factor, since $|\ltthre| \sim Q \sim \Delta$ and $P_\perp^2\sim \omega \ltthre^2 \sim z Q^2$. Fortunately, we can perform the auxiliary $\ltthre$ integral analytically, such that
\begin{align}
    &\int\frac{\der^2\ltthre}{(2\pi)} \Hcal_{\rm R1\times R1^*}^j \Hcal_{\rm R1\times R1^*}^{j*}=\frac{\bar Q^2}{2z_g(1-z_g)}\left[\frac{1}{P_\perp^2\bar Q^2}+\frac{1}{P_\perp^2(P_\perp^2+\bar Q^2)}-\frac{2\ln(1+P_\perp^2/\bar Q^2)}{\Pt^4}+\frac{1}{(\Pt-\qt)^2\bar Q^2}\right.\nonumber\\
    &+\frac{1}{(\Pt-\qt)^2((\Pt-\qt)^2+\bar Q^2)}-\frac{2\ln(1+(\Pt-\qt)^2/\bar Q^2)}{(\Pt-\qt)^4}-2\frac{\Pt\cdot(\Pt-\qt)}{\Pt^4(\Pt-\qt)^4(P_\perp^2-(\Pt-\qt)^2)}\nonumber\\
    &\left.\times \left(\frac{(P_\perp^2-(\Pt-\qt)^2)(\Pt-\qt)^2P_\perp^2}{\bar Q^2}+(\Pt-\qt)^4\ln(1+P_\perp^2/\bar Q^2)-\Pt^4\ln(1+(\Pt-\qt)^2/\bar Q^2)\right)\right]\,,\label{eq:l3-int}
\end{align}
with $\bar Q^2=zQ^2$. The integral over $z_g$ in Eq.\,\eqref{eq:FUUL-R1R1-app1} becomes trivial since $z_g$ does not appear inside the square bracket of Eq.\,\eqref{eq:l3-int}:
\begin{align}
    \int_0^1\der z_g (1-z_g)=\frac{1}{2}\,.
\end{align}
The last step is to do the $z$ integral and to take the limit $Q\gg P_\perp$ of the result. These two manipulations can be performed together thanks to the change of variable $z\to u=zQ^2/P_\perp^2$, such that the integral over $u$ goes from $0$ to $Q^2/P_\perp^2\to \infty$ in the limit $Q^2\gg P_\perp^2$:
\begin{align}
     F^{\rm R1\times R1^*}_{UU,L}&=\frac{\alpha_sC_FN_c}{8\pi^5} \sum_{i=q} e_i^2  \int_{\Bt,\qt}\mathcal{D}_F(x,\qt)\int_0^\infty\der u \ 2u \left[\frac{1}{u}+\frac{1}{1+u}-2\ln\left(1+\frac{1}{u}\right)-\frac{\Pt\cdot(\Pt-\qt)}{(P_\perp^2-(\Pt-\qt)^2)}\right.\nonumber\\
    &\left.\times\left(\frac{P_\perp^2-(\Pt-\qt)^2}{(\Pt-\qt)^2u}+\ln\left(1+\frac{1}{u}\right)-\frac{\Pt^4}{(\Pt-\qt)^4}\ln\left(1+\frac{(\Pt-\qt)^2}{P_\perp^2u}\right)\right)\right] \\
&=\frac{\alpha_s C_F}{2\pi} \sum_{i=q} e_i^2 \times \frac{N_c}{4\pi^4}\int_{\Bt,\qt}\mathcal{D}_F(x,\qt)\left[1-\frac{\Pt\cdot(\Pt-\qt)\ln(P_\perp^2/(\Pt-\qt)^2)}{(P_\perp^2-(\Pt-\qt)^2)}\right]\,,
\end{align}
which is exactly the result given by Eq.\,(9) in the main text. The integral over $u$ is controlled by $u\lesssim 1$, meaning that the $z$ integration is dominated by the physical endpoint $z\sim P_\perp^2/Q^2$ as discussed in the main text. It is remarkable that the integral over $u$ coming from the complicated momentum space hard factor in Eq.\,\eqref{eq:l3-int} eventually gives the same $\Pt$ dependence as for the sea quark TMD involved in the transversely polarised photon case at LO.

It is straightforward to extend these results to the jet cross-section differential both in $\Pt$ and $\xi\equiv  k^-/P^-$, at LO. If one does not integrate over $z$, one must trade the $z$-dependence of the leading power cross-section to its $\xi$-dependence, using the onshell relation for the measured quark in the target fragmentation region, which reads $zQ^2=x\Pt^2/\xi$. The result is
\begin{align}
    F_{UU,L}(x,\xi,\Pt)&=\frac{\alpha_s C_F}{2\pi} \sum_{i=q} e_i^2 x\mathcal{F}_q(x,\xi,\Pt)\,,
\end{align}
with
\begin{align}
    x\mathcal{F}_q(x,\xi,\Pt)\equiv &\frac{N_c}{8\pi^4} \frac{x}{(x+\xi)^2}\int_{\bt,\qt}\mathcal{D}_F\left(x+\xi,\qt\right)\left\{1+\frac{(x+\xi)^2\Pt^2(\Pt-\qt)^2}{[\xi(\Pt-\qt)^2+x\Pt^2]^2}-\frac{2(\xi+x)[\Pt\cdot(\Pt-\qt)]}{[\xi(\Pt-\qt)^2+x\Pt^2]}\right\} \,.
\end{align}
It is easy to check that if one integrates $\xi$ between 0 and $1$ in this expression (neglecting the mild logarithmic $\xi$-dependence inside $\mathcal{D}_F$) and expands for $x\ll 1$ (alternatively, one can simply integrate $\xi$ between 0 and $\infty$), one recovers the expression for $x\mathcal{F}_q(x,\Pt)$ in the Letter.

\section{C. Proof of Eqs.\,(11)-(12) in the main text}

We now turn to the derivation of the leading power contribution coming from a gluon jet measurement in single-inclusive jet production in DIS mediated by a longitudinally polarised virtual photon. The extraction of the LP component out of the coordinate space expressions obtained in \cite{Caucal:2024cdq} is more complicated than in the quark jet case. To cross-check our final result presented in the last subsection and given by Eq.\,(12) in the main text, we have performed an independent calculation based on the momentum space expression of the cross-section for $\gamma_L^*+A\to q\bar q+  $gluon with $z_g\sim P_\perp^2/Q^2\ll 1$ production, in which the $q\bar q$ pair is subsequently integrated out.

In the following, we divide the calculation into three contributions according to the notations of \cite{Caucal:2024cdq}: 

(i) the NLO-4 term coming from Feynman graphs where the gluon is emitted before the shockwave both in the amplitude and in the complex conjugate amplitude (cf.~bottom-left diagram in Fig.\,(1) of the Letter), 

(ii) the NLO-1 term coming from interferences between gluon emission before and after the shockwave and,

(iii) the NLO-0+NLO-3 terms coming from diagrams where the gluon is emitted after the shockwave both in the amplitude and complex conjugate amplitude (cf.~bottom-right diagram in Fig.\,(1) of the Letter).

\subsection{The NLO-4 contribution}

The contribution of Eq.\,(3.73) in \cite{Caucal:2024cdq} to $F_{UU,L}$ reads
\begin{align}
    F_{UU,L}^{\rm NLO4}&=\frac{\alpha_s N_c}{4\pi^8} \sum_{i=q} e_i^2  \int_{\xt,\yt,\zt,\zt'} \!\!\!\!\!\!\!\!\!\!\!\! e^{-i\Pt\cdot\rzzpt}\int_0^1\frac{\der z}{z}\int_0^{1-z}\der z_1 z_1^2(1-z_1-z)^2Q^4K_0(QX_R)K_0(QX_R')\nonumber\\
    &\times \mathfrak{Re}\left[\Xi_{\rm NLO4,g}(\xt,\yt,\zt,\zt')\right]\left[\left(1+\frac{z}{z_1}+\frac{z^2}{2z_1^2}\right)\frac{\rzxt\cdot\rzpxt}{\rzxt^2\rzpxt^2}-\left(1+\frac{z}{2z_1}+\frac{z}{2(1-z_1-z)}\right)\frac{\rzyt\cdot\rzpxt}{\rzyt^2\rzpxt^2}\right]\,,\label{eq:FUUL-g-NLO4}
\end{align}
with the CGC correlator given by
\begin{align}
    \Xi_{\rm NLO4,g}(\xt,\yt,\zt,\zt')&=\frac{N_c}{2}\left\langle 1-D_{xz}D_{zy}-D_{yz'}D_{z'x}+D_{z'z}D_{zz'}\right\rangle_x-\frac{1}{2N_c}\left\langle2-D_{xy}-D_{yx}\right\rangle\,,
\end{align}
and the effective size of the $q\bar qg$ system 
\begin{align}
X_R&=z_1(1-z_1-z)\rxyt^2+z_1z\rzxt^2+(1-z_1-z)z\rzyt^2\,,\\
X_R'&=z_1(1-z_1-z)\rxyt^2+z_1z\rzpxt^2+(1-z_1-z)z\rzpyt^2\,.
\end{align}
Note that since the gluon-jet is tagged, we have $z_g=z$ and we integrate over the longitudinal momentum fraction $z_1$ of the quark. 

We first do a change of transverse coordinates using $\Bt=z_1\xt+(1-z_1)\yt$ and $\rxyt=\xt-\yt$. Note that since $z\ll 1$, $\Bt$ physically corresponds to the impact parameter of the virtual photon. We then implement the approximations discussed in the main text, namely (i) the integral over $z$ is dominated by $z\sim P_\perp^2/Q^2\ll 1$ when $Q^2\gg P_\perp^2$, (ii) the integral over transverse coordinates is controlled by dipole sizes such that $r_{xy}\sim 1/Q$ and $r_{zB}\sim r_{z'B}\sim 1/P_\perp$.

In this limit, we have $X_R\approx z_1(1-z_1)\rxyt^2+z\rzbt^2$ and $X_R'\approx z_1(1-z_1)\rxyt^2+z\rzpbt^2$. The next step is to expand $\Xi_{\rm NLO4,g}$ in the regime $r_{xy}\ll r_{zB},r_{z'B}$:
\begin{align}
    &\mathfrak{Re}(\Xi_{\rm NLO4,g}(\xt,\yt,\zt,\zt'))=\frac{N_c}{2}\mathfrak{Re}\left\langle 1-D_{Bz}D_{zB}-D_{Bz'}D_{z'B}+D_{zz'}D_{z'z}\right\rangle\nonumber\\&-(1-2z_1)\rxyt^j \frac{N_c}{2} \mathfrak{Re} \left\langle\frac{1}{N_c^2}\Tr(V_BV_z^\dagger)\Tr(V_z\partial^j V_B^\dagger)\right\rangle-(1-2z_1)\rxyt^j \frac{N_c}{2} \mathfrak{Re} \left\langle\frac{1}{N_c^2}\Tr(\partial^j V_BV_{z'}^\dagger)\Tr(V_{z'}V^\dagger_B)\right\rangle+\mathcal{O} (\rxyt^2)\,, \label{eq:NLO4-exp}
\end{align}
where the terms of order $\rxyt^2$ can be neglected as they would necessarily yield $1/Q^2$ suppressed contributions after forming the product with the quantity inside the square bracket in Eq.\,\eqref{eq:FUUL-g-NLO4}.

Indeed, this quantity must be similarly expanded up to the order $\mathcal{O}(\rxyt^2)$ or $\mathcal{O}(z)$, as
\begin{align}
    &\left[\left(1+\frac{z}{z_1}+\frac{z^2}{2z_1^2}\right)\frac{\rzxt\cdot\rzpxt}{\rzxt^2\rzpxt^2}-\left(1+\frac{z}{2z_1}+\frac{z}{2(1-z_1-z)}\right)\frac{\rzyt\cdot\rzpxt}{\rzyt^2\rzpxt^2}\right]\approx-\frac{\rzpbt^j\rxyt^k}{\rzpbt^2\rzbt^2}\left[\delta^{jk}-\frac{2\rzbt^k\rzbt^l}{\rzbt^2}\right]\nonumber\\
    &+\frac{z(1-2z_1)}{2z_1(1-z_1)}\frac{\rzbt\cdot\rzpbt}{\rzbt^2\rzpbt^2}+(1-z_1)\frac{\rxyt^k\rxyt^l}{\rzbt^2\rzpbt^2}\left[\delta^{jk}-\frac{2\rzbt^j\rzbt^k}{\rzbt^2}\right]\left[\delta^{jl}-\frac{2\rzpbt^j\rzpbt^l}{\rzpbt^2}\right]+\mathcal{O}(z^2,r_{xy}^3)\,.\label{eq:WWkernel-exp}
\end{align}
In order to get a leading power contribution, one should consider the terms in the product between Eq.\,\eqref{eq:NLO4-exp} and Eq.\,\eqref{eq:WWkernel-exp} which are at most $\mathcal{O}(z^0r_{xy}^2)$ or $\mathcal{O}(z^1 r_{xy}^0)$. Among those terms, the only one which is not proportional to $1-2z_1$ is 
\begin{align}
    \frac{N_c}{2}\mathfrak{Re}\left\langle 1-D_{Bz}D_{zB}-D_{Bz'}D_{z'B}+D_{zz'}D_{z'z}\right\rangle\times (1-z_1)\frac{\rxyt^k\rxyt^l}{\rzbt^2\rzpbt^2}\left[\delta^{jk}-\frac{2\rzbt^j\rzbt^k}{\rzbt^2}\right]\left[\delta^{jl}-\frac{2\rzpbt^j\rzpbt^l}{\rzpbt^2}\right]\,.\label{eq:nv-term}
\end{align}
It is easy to check that the terms proportional to $1-2z_1$ will cancel after integrating over $z_1$, since 
\begin{align}
    \int_0^1\der z_1 (1-2z_1) = 0\,.
\end{align}

The CGC correlator in Eq.\,\eqref{eq:nv-term} can be fully written in term of the gluon-gluon dipole $D^g_{xz}$,
\begin{align}
    D^g_{Bz}&\equiv\frac{1}{N_c^2-1}U_{\Bt}^{ab}U_{\zt}^{\dagger,ba}\,,
\end{align}
by mean of the Fierz identity, such that
\begin{align}
    D_{Bz}D_{zB}&=\frac{N_c^2-1}{N_c^2}D_{Bz}^g+\frac{1}{N_c^2}\,.
\end{align}
Using this equation, we find that the CGC correlator in Eq.\,\eqref{eq:nv-term} reduces to a sum of gluon-gluon dipoles as discussed in the main text:
\begin{align}
  \frac{N_c}{2}\mathfrak{Re}\left\langle 1-D_{Bz}D_{zB}-D_{Bz'}D_{z'B}+D_{zz'}D_{z'z}\right\rangle&=C_F\left\langle 1-D^g_{zB}-D^g_{Bz'}+D^g_{zz'}\right\rangle_x\,,\\    &=C_F\int_{\qt}\left(e^{i\qt\cdot\rzbt}-1\right)\left(e^{-i\qt\cdot\rzpbt}-1\right)\mathcal{D}_A(x,\qt)\,.
\end{align}

Combining all these results together, the leading power contribution of Eq.\,\eqref{eq:FUUL-g-NLO4} can be expressed as
\begin{align}
    F_{UU,L}^{\rm NLO4,LP}&=\frac{\alpha_s N_c}{4\pi^8} \sum_{i=q} e_i^2  \int_{\Bt,\rxyt,\rzbt,\rzpbt} \!\!\!\!\!\!\!\!\!\!\!\! e^{-i\Pt\cdot\rzzpt}C_F \mathfrak{Re} \int_{\qt}\left(e^{i\qt\cdot\rzbt}-1\right)\left(e^{-i\qt\cdot\rzpbt}-1\right)\mathcal{D}_A(x,\qt)\nonumber\\
    &\times\int_0^1\frac{\der z}{z}\int_0^{1}\der z_1 z_1^2(1-z_1)^3Q^4K_0(QX_R)K_0(QX_R')\frac{\rxyt^k\rxyt^l}{\rzbt^2\rzpbt^2}\left[\delta^{jk}-\frac{2\rzbt^j\rzbt^k}{\rzbt^2}\right]\left[\delta^{jl}-\frac{2\rzpbt^j\rzpbt^l}{\rzpbt^2}\right] \\
    &=\frac{2\alpha_sC_F N_cQ^2}{\pi^5} \sum_{i=q} e_i^2  \int_0^1\frac{\der z}{z}\int_0^{1}\der z_1\int_{\Bt,\qt}\mathcal{D}_A(x,\qt) z_1(1-z_1)^2\int\frac{\der^2\ltthre}{(2\pi)}\Hcal^j_{\rm NLO4}\Hcal_{\rm NLO4}^{j*}\,,
\end{align}
with 
\begin{align}
    \mathcal{H}^{j}_{\rm NLO4}&=\int\frac{\der^2\rzbt}{(2\pi)}\int\frac{\der^2\rxyt}{(2\pi)} e^{-i\Pt\cdot\rzbt}\left(e^{i\qt\rzbt}-1\right)e^{i\ltthre\cdot\rxyt}\frac{\bar Q_1\rxyt^k}{\rzbt^2}\left[\delta^{jk}-2\frac{\rzbt^j\rzbt^k}{\rzbt^2}\right]K_0\left(\bar Q_1\sqrt{\rxyt^2+\omega\rzbt^2}\right)\,,\\
    &=\frac{-i\bar Q_1(\Pt-\qt)^2\ltthre^k}{(\ltthre^2+\bar Q_1^2)^2((\Pt-\qt)^2+\omega(\ltthre^2+\bar Q_1^2))}\left[\delta^{jk}-\frac{2(\Pt-\qt)^j(\Pt-\qt)^k}{(\Pt-\qt)^2}\right]\nonumber\\
    &+\frac{i\bar Q_1\Pt^2\ltthre^k}{(\ltthre^2+\bar Q_1^2)^2(\Pt^2+\omega(\ltthre^2+\bar Q_1^2))}\left[\delta^{jk}-\frac{2\Pt^j\Pt^k}{\Pt^2}\right]\,,
\end{align}
where we have introduced the shorthand variables $\bar Q_1^2=z_1(1-z_1)Q^2$ and $\omega=z/(z_1(1-z_1))$. To get the second equality, we have used similar tricks as in section B of this Supplemental Material.

Integrating over $\ltthre$, $z_1$ and $z$ between some lower cut-off $z_0$ and $1$ yields
\begin{align}
    &F_{UU,L}^{\rm NLO4,LP}=\frac{\alpha_s C_FN_c}{6\pi^5} \sum_{i=q} e_i^2 \int_{\Bt,\qt}\!\!\!\!\!\!\!\!\!\!\! \mathcal{D}_A(x,\qt)\left[-\frac{11}{6}+\frac{1}{2}\ln\left(\frac{\Pt^2}{z_0Q^2}\right)+\frac{1}{2}\ln\left(\frac{(\Pt-\qt)^2}{z_0Q^2}\right)+\left(1-\frac{2(\Pt\cdot(\Pt-\qt))^2}{\Pt^2(\Pt-\qt)^2}\right)\right.\nonumber\\
    &\left.\times\left(-\frac{5}{6}+\frac{\Pt^2+(\Pt-\qt)^2}{2(\Pt^2-(\Pt-\qt)^2)}\ln\left(\frac{(\Pt-\qt)^2}{\Pt^2}\right)+\frac{1}{2}\ln\left(\frac{\Pt^2}{z_0Q^2}\right)+\frac{1}{2}\ln\left(\frac{(\Pt-\qt)^2}{z_0Q^2}\right)\right)\right]\,.
\end{align}
The dependence upon $z_0$ and $Q^2$ in this expression will cancel after adding the NLO-1 contribution.

\subsection{The NLO-1 contribution}

Let us now consider the contribution of Eq.\,(3.72) in \cite{Caucal:2024cdq} to $F_{UU,L}$:
\begin{align}
     F_{UU,L}^{\rm NLO1} &=\frac{\alpha_sN_c}{4\pi^8}\sum_{i=q} e_i^2 \int_{\xt,\yt,\zt,\zt'} \!\!\!\!\!\!\!\!\!\!\!\!\!\!\!\!\!\!\!\!\!\!\!\! e^{-i\Pt \cdot \rzzpt}\int_0^1\frac{\der z}{z}\int_0^{1-z}\der z_1 \ e^{i\frac{z_1}{z}\Pt\cdot\rzpxt}  z_1^2\left (1-z_1-z\right)^2\left(1+\frac{z_1}{z}\right)Q^4K_0(\bar{Q}_{\rm R2}r_{z'y})K_0(QX_{R})  \nonumber\\
    &\times \mathfrak{Re}\left[\Xi_{\rm NLO,1}(\xt,\yt,\zt,\zt')\right]\left[\left(1+\frac{z}{2z_1}+\frac{z}{2(1-z_1-z)} \right)\frac{\rzyt\cdot \rzpxt}{\rzyt^2 \rzpxt^2}- \left(1+\frac{z}{z_1}+\frac{z^2}{2z_1^2} \right)\frac{\rzxt\cdot \rzpxt}{\rzxt^2 \rzpxt^2}\right] +c.c.\,, \label{eq:gjet-NLO1}
\end{align}
with $\bar Q^2_{\rm R2}=(1-z_1-z)(z_1+z)Q^2$ and the CGC correlator
\begin{align}
     \Xi_{\rm NLO,1}(\xt,\yt,\zt,\zt')&=\frac{N_c}{2}\left\langle 1-D_{yz'}-D_{xz}D_{zy}+D_{zz'}D_{xz}\right\rangle-\frac{1}{2N_c}\left\langle 1-D_{xy}-D_{yz'}+D_{xz'}\right\rangle\,.
\end{align}
As for the NLO-4 term, we aim at extracting the LP contribution out of this expression. We also change the variables to $\Bt=z_1\xt+(1-z_1)\yt$ and $\rxyt=\xt-\yt$.

We look for a gluon jet in the target fragmentation region with $z\sim \Pt^2/Q^2$. Defining $\Lt=z_1\Pt/z$ such that $L_\perp\sim Q^2/P_\perp$, we have the hierarchy $L_\perp \gg Q \gg P_\perp $ such that $r_{z'x}\ll r_{xy}\ll r_{zB}$ from the phases. With this hierarchy of transverse distances in mind, we first simplify the color structure using $\zt'=\xt$
\begin{align}
    \Xi_{\rm NLO,1}(\xt,\yt,\zt,\zt')&\approx \frac{N_c}{2}\left\langle 1-D_{yx}-D_{xz}D_{zy}+D_{zx}D_{xz}\right\rangle-\frac{1}{2N_c}\left\langle 2-D_{xy}-D_{yx}\right\rangle\,,
\end{align}
and we further expand it around $\Bt$ up to the first order in $\rxyt$ as for the NLO-4 color structure in the previous subsection:
\begin{align}
   \mathfrak{Re}(\Xi_{\rm NLO,1}(\xt,\yt,\zt,\zt'))&=\rxyt^j\frac{N_c}{2} \mathfrak{Re} \left\langle\frac{1}{N_c^2}\Tr(V_BV_z^\dagger)\Tr(V_z\partial^j V_B^\dagger)\right\rangle+\mathcal{O}(\rxyt^2) \\
   &=\rxyt^j \frac{C_F}{2}\int\der^2\qt e^{-i\qt\cdot\rzbt}(i\qt^j )\mathcal{D}_A(x,\qt)\,.
\end{align}
The quantity inside the big square bracket in Eq.\,\,\eqref{eq:gjet-NLO1} is identical, up to an overall minus sign, to Eq.\,\eqref{eq:WWkernel-exp}. However, since the expansion of the CGC correlator starts at the order $\mathcal{O}(r_{xy})$, it is sufficient to truncate Eq.\,\eqref{eq:WWkernel-exp} up to the order one in $\mathcal{O}(r_{xy})$ and the zeroth order in $z$. We thus get
\begin{align}
     F_{UU,L}^{\rm NLO1, LP} &=\frac{\alpha_sN_c}{8\pi^7} \sum_{i=q} e_i^2 \int_{\Bt,\rxyt,\zt,\zt'} \!\!\!\!\!\! e^{-i\Pt \cdot \rzbt}C_F\int_{\qt} e^{-i\qt\cdot\rzbt}(i\qt^k)\mathcal{D}_A(x,\qt)\nonumber\\
     &\times \int_0^1\frac{\der z}{z^2}\int_0^{1}\der z_1 \ e^{i\frac{z_1}{z}\Pt\cdot\rzpbt}  z_1^3\left (1-z_1\right)^2Q^4K_0(\bar{Q}_{\rm R2}r_{zB})\rxyt^k K_0(QX_{R})\frac{\rzpbt^l\rxyt^j}{\rzpbt^2\rzbt^2}\left[\delta^{lj}-\frac{2\rzbt^l\rzbt^j}{\rzbt^2}\right]   +c.c. \nonumber\\
     &=\frac{\alpha_s N_c C_F Q^2}{\pi^5} \sum_{i=q} e_i^2 \int_0^1\frac{\der z}{z}\int_0^{1}\der z_1 \  \int_{\Bt,\qt} \mathcal{D}_A(x,\qt) z_1(1-z_1)\frac{(-\qt^k\Pt^l)}{\Pt^2}\Hcal_{\rm NLO1}^{lk} + c.c.\,,
\end{align}
with
\begin{align}
    \Hcal_{\rm NLO1}^{lk}&=\int\frac{\der^2\rxyt}{(2\pi)}\frac{\der^2\rzbt}{(2\pi)}  e^{-i(\Pt+\qt)\cdot \rzbt} \frac{\rxyt^k\rxyt^j}{\rzbt^2}\left[\delta^{lj}-2\frac{\rzbt^l\rzbt^j}{\rzbt^2}\right]\bar Q_1^2K_0(\bar{Q}_{1}r_{xy})K_0\left(\bar Q_1\sqrt{\rxyt^2+\omega\rzbt^2}\right) \\
    &=\frac{1}{12}\left(\delta^{lk}-\frac{2(\Pt+\qt)^l(\Pt+\qt)^k}{(\Pt+\qt)^2}\right)\left[-\frac{1}{\bar Q_1^2}+\frac{3\omega}{(\Pt+\qt)^2}+\frac{6\omega^2\bar Q_1^2}{(\Pt+\qt)^4}\right.\nonumber\\
    &\left.-\frac{6\omega^2\bar Q_1^2((\Pt+\qt)^2+\omega\bar Q_1^2)}{(\Pt+\qt)^6}\ln\left(1+\frac{(\Pt+\qt)^2}{\omega\bar Q_1^2}\right)\right]\,.
\end{align}
Integrating over $z_1$ and $z$ between $z_0$ and $1$ yields (we also make the change of variable $\qt\to -\qt$)
\begin{align}
    F_{UU,L}^{\rm NLO1, LP} &=\frac{\alpha_s N_c C_F}{6\pi^5} \sum_{i=q} e_i^2 \int_{\Bt,\qt}\mathcal{D}_A(x,\qt)\frac{\qt^k\Pt^i}{\Pt^2}\left(\delta^{ik}-\frac{2(\Pt-\qt)^i(\Pt-\qt)^k}{(\Pt-\qt)^2}\right)\left[\frac{5}{6}-\ln\left(\frac{(\Pt-\qt)^2}{z_0Q^2}\right)\right] \\
    &=\frac{\alpha_s N_c C_F}{6\pi^5} \sum_{i=q} e_i^2 \int_{\Bt,\qt}\mathcal{D}_A(x,\qt)\left(1-\frac{2(\Pt\cdot(\Pt-\qt))^2}{\Pt^2(\Pt-\qt)^2}+\frac{\Pt\cdot(\Pt-\qt)}{\Pt^2}\right)\left[\frac{5}{6}-\ln\left(\frac{(\Pt-\qt)^2}{z_0Q^2}\right)\right]\,.
\end{align}

\subsection{The NLO-0+NLO-3 contributions}

We finally consider the diagrams where the gluon sourcing the measured jet does not interact with the shockwave. At the cross-section level, they contribute to $F_{UU,L}$ as (see Eqs.\,(3.70)-(3.71) in \cite{Caucal:2024cdq})
\begin{align}
     F_{UU,L}^{\rm NLO0}&=\frac{(-\alpha_s)C_FN_cQ^4}{4\pi^7}\sum_{i=q} e_i^2 \int_{\xt,\yt,\xt'} e^{-i\Pt\cdot\rxxtp}\int_0^1\frac{\der z}{z}\int_0^{1-z}\frac{\der z_1}{z^2} \ e^{-i\frac{z_1}{z}\Pt \cdot \rxxtp}K_0(\bar Q_{\mathrm{R}2}r_{x'y})K_0(\bar Q_{\mathrm{R}2}r_{xy})\nonumber\\
    &\times \mathfrak{Re}\left[\Xi_{\rm LO}(\xt,\yt,\xt')\right ]\left(1-z_1-z\right)^2\left(z_1+z\right)^2\left(z_1(z_1+z)+\frac{z^2}{2}\right)\ln\left(\frac{Q^2\rxxtp^2R^2z_1^2}{c_0^2}\right)\,,  \label{eq:gjet-NLO0}
\end{align}
for the NLO-0 contribution adapted with the jet definition of \cite{Caucal:2024vbv} and
\begin{align}
   F_{UU,L}^{\rm NLO3}&=\frac{\alpha_sN_cQ^4}{4\pi^8}\sum_{i=q} e_i^2 \int_{\xt,\xt',\yt,\yt'}e^{-i\Pt\cdot\rxyt}\int_0^1\frac{\der z}{z}\int_0^{1-z}\frac{\der z_1}{z^2}e^{-i\frac{\Pt}{z}\cdot\left((1-z_1)\ryytp+z_1\rxxtp\right)} K_0(\bar Q_1 r_{xy})K_0(\bar Q_{\rm R2}r_{x'y'})\ \nonumber\\
    &\times \mathfrak{Re}\left[\Xi_{\rm NLO,3}(\xt,\yt;\xt',\yt')\right]z_1(1-z_1)(z_1+z)(1-z_1-z)\left(z_1(1-z_1-z)+\frac{z(1-z)}{2}\right)\frac{\ryytp\cdot\rxxtp}{\ryytp^2\rxxtp^2}\,, \label{eq:gjet-NLO3}
\end{align}
for the NLO-3 one. For this jet definition depending on the radius parameter $R$ (assumed to be $\ll 1$), two particles are clustered within the same jet provided $M_{ij}^2/(z_iz_jQ^2)\le R^2$, with $M_{ij}$ the invariant mass of the pair and~$z_i=k_i\cdot P/(P\cdot q)$. In practice, it amounts to changing the argument in the logarithm in Eq.\,\eqref{eq:gjet-NLO0} from $P_\perp^2\rxxtp^2R^2 z_1^2/(z_g^2c_0^2)$ as shown in \cite{Caucal:2024cdq} to $Q^2\rxxtp^2R^2 z_1^2/c_0^2$ where $c_0=2e^{-\gamma_E}$. As we shall see, the LP gluon jet contribution does not depend on $R$ at our perturbative order, meaning that it is not sensitive to the singularity arising when the integrated quark becomes collinear to the measured gluon. Nevertheless, we expect the jet definition proposed in \cite{Caucal:2024vbv} to play an important role at NNLO in the CGC, as observed in the transversely polarised photon case at NLO where it guarantees that TMD factorisation both in the current and target fragmentation region is preserved by quantum corrections.

In these expressions Eqs.\,\eqref{eq:gjet-NLO0}-\eqref{eq:gjet-NLO3}, the CGC correlators are defined as
\begin{align}
    \Xi_{\rm LO}(\xt,\yt,\xt')&=\left\langle D_{xx'}-D_{xy}-D_{yx'}+1\right\rangle\,,\\
    \Xi_{\rm NLO,3}(\xt,\yt,\xt',\yt')&=\frac{N_c}{2}\left\langle 1-D_{xy}-D_{y'x'}+D_{xy}D_{y'x'}\right\rangle-\frac{1}{2N_c}\left\langle 1-D_{xy}-D_{y'x'}+Q_{xy;y'x'}\right\rangle\,.
\end{align}
with the quadrupole $Q_{xy;y'x'}=\frac{1}{N_c}\Tr\left[V(\xt)V^\dagger(\yt)V(\yt')V^\dagger(\xt')\right]$.
The leading power term of these two expressions can be obtained by considering the regime $z\sim P_\perp^2/Q^2$ such that the phases 
\begin{align}
    e^{-i\frac{z_1}{z}\Pt \cdot \rxxtp}\,,\quad e^{-i\frac{\Pt}{z}\cdot\left((1-z_1)\ryytp+z_1\rxxtp\right)}\,,
\end{align}
dominate and therefore impose $\rxxtp=0$, $\ryytp=0$, i.e.~$\xt=\xt'$ and $\yt=\yt'$. With theses identifications, the CGC correlators simplify 
\begin{align}
    \mathfrak{Re}\left[\Xi_{\rm LO}\right]&=2\mathfrak{Re}(1-\left\langle D_{xy}\right\rangle) \\
    &=2\mathfrak{Re}\int\der^2\qt\left(1-e^{i\qt\cdot\rxyt}\right)\mathcal{D}_F(x,\qt)\,,\\
    \mathfrak{Re}\left[\Xi_{\rm NLO3}\right]&=2C_F \mathfrak{Re}(1-D_{xy})+\frac{N_c}{2}\mathfrak{Re}(\left\langle D_{xy}D_{yx}\right\rangle-1) \\
    &=2C_F\mathfrak{Re}\int\der^2\qt\left(1-e^{i\qt\cdot\rxyt}\right)\mathcal{D}_F(x,\qt)+C_F\mathfrak{Re}\int\der^2\qt\left(e^{i\qt\cdot\rxyt}-1\right)\mathcal{D}_A(x,\qt)\,.
\end{align}
Let us first consider the contribution depending on the gluon-gluon dipole $\mathcal{D}_A(x,\qt)$. The latter only comes from $F_{UU,L}^{\rm NLO3}$. Using
\begin{align}
    \int\frac{\der^2\rxxtp}{(2\pi)}\int\frac{\der^2\ryytp}{(2\pi)}e^{-i\frac{\Pt}{z}\cdot\left((1-z_1)\ryytp+z_1\rxxtp\right)}\frac{\ryytp\cdot\rxxtp}{\ryytp^2\rxxtp^2}&=-\frac{z^2}{z_1(1-z_1)}\frac{1}{\Pt^2}\,,
\end{align}
we get
\begin{align}
    \left.F_{UU,L}^{\rm NLO3}\right|_A&\approx \frac{\alpha_sC_FN_cQ^4}{\pi^6\Pt^2}\sum_{i=q} e_i^2 \int_{\Bt,\qt} \mathcal{D}_A(x,\qt)\int\der^2\rxyt\left(1-e^{i\qt\cdot\rxyt}\right)e^{-i\Pt\cdot\rxyt}\int_{z_0}^{P_\perp^2/Q^2}\frac{\der z}{z}\int_0^{1}\der z_1 z_1^2(1-z_1)^2K_0^2(\bar Q_1 r_{xy})\,.
\end{align}
The upper boundary of the $z$ integral follow from the approximation $z\lesssim P_\perp^2/Q^2$ used to simplify the expressions for the NLO,0 and NLO,3 terms. The lower cut-off $z_0$ regulates the $z\to0$ logarithmic divergence.

For the contribution depending on the $q\bar q$ dipole $\mathcal{D}_F(x,\qt)$, coming both from NLO-0 and NLO-3, we have
\begin{align}
    \left.F_{UU,L}^{\rm NLO3}\right|_F+\left.F_{UU,L}^{\rm NLO0}\right|_F&\approx \frac{2\alpha_sC_FN_cQ^4}{\pi^6\Pt^2}\sum_{i=q} e_i^2 \int_{\Bt,\qt} \mathcal{D}_F(x,\qt)\int\der^2\rxyt\left(1-e^{i\qt\cdot\rxyt}\right)\left(1-e^{-i\Pt\cdot\rxyt}\right)\nonumber\\
    &\times \int_{z_0}^{P_\perp^2/Q^2}\frac{\der z}{z}\int_0^{1}\der z_1 z_1^2(1-z_1)^2K_0^2(\bar Q_1 r_{xy})\,.
\end{align}
The $K_0$ Bessel function imposes that $r_{xy}\lesssim 1/\bar Q_1\ll 1/P_\perp,1/q_\perp$. Therefore, one can expand the remaining phases up to second order in $r_{xy}$. By rotational invariance, the piece depending on $\mathcal{D}_F(x,\qt)$ vanishes since it depends on $\qt\cdot\Pt$ only. For the piece depending on $\mathcal{D}_A(x,\qt)$, we use 
\begin{align}
    \int\frac{\der^2\rxyt}{(2\pi)}\rxyt^2 K_0^2(\bar Q_1r_{xy})&=\frac{1}{3\bar Q_1^4}\,,
\end{align}
and we eventually get
\begin{align}
    F_{UU,L}^{\rm NLO0+3, LP}&=\frac{\alpha_sN_cC_F}{6\pi^5}\sum_{i=q} e_i^2 \frac{1}{P_\perp^2}\int_{z_0}^{P_\perp^2/Q^2}\frac{\der z}{z}\int_{\Bt,\qt}q_\perp^2\mathcal{D}_A(x,\qt)\,.\label{eq:NLO0-3-LP}
\end{align}
This LP contribution suffers several logarithmic divergences which are regulated by physical scales. For $z\to 0$, we have introduced a lower cut-off $z_0$, which is physically determined by imposing the gluon being on-shell, $2k_g^+k_g^-=P_\perp^2$, and the condition $k_g^- \le P^-$ from minus momentum conservation. Resolving this inequality with $k_g^+=z q^+$ yields $z\ge x_{\rm Bj}P_\perp^2/Q^2$. We shall then use $z_0=xP_\perp^2/Q^2$. The integral over $\qt$ is also divergence for $q_\perp\to\infty$, since typically $\mathcal{D}_A(x,\qt)\propto 1/\qt^4$ at large $q_\perp$. Our derivation sets the natural scale $Q^2$ at which one should cut the $\qt$ integration, since we have performed the expansion for $1/\bar Q_1\ll 1/q_\perp$.

\subsection{Final result for the NLO gluon-jet contribution at leading power}

Adding all leading power contributions, we get
\begin{align}
  F_{UU,L}^{\textrm{NLO,LP-}g}&= \frac{\alpha_s C_FN_c}{6\pi^5}\sum_{i=q} e_i^2 \int_{\Bt,\qt}\mathcal{D}_A(x,\qt)\left\{\ln\left(\frac{1}{x}\right)\frac{\qt^2}{\Pt^2}-1-\frac{(\Pt-\qt)^2-\qt^2}{2\Pt^2}\ln\left(\frac{(\Pt-\qt)^2}{\Pt^2}\right)\right.\nonumber\\
&\left.+\left(1-\frac{2(\Pt\cdot(\Pt-\qt))^2}{\Pt^2(\Pt-\qt)^2}\right)\frac{(\Pt-\qt)^2}{\Pt^2-(\Pt-\qt)^2}\ln\left(\frac{(\Pt-\qt)^2}{\Pt^2}\right)\right\}\,.
\end{align}
The explicit logarithmic dependence upon $x$ fully comes from Eq.\,\eqref{eq:NLO0-3-LP}, where $z_0=xP_\perp^2/Q^2$ has been used.
 In the dilute limit $P_\perp\gg q_\perp$, we have
\begin{align}
   F_{UU,L}^{\textrm{NLO,LP-}g} &= \frac{\alpha_s C_F}{6\pi^5}\sum_{i=q} e_i^2 \left[N_c\ln\left(\frac{1}{x}\right)+\frac{(-11)N_c}{6}\right]\frac{1}{\Pt^2}\int_{\Bt,\qt}\qt^2\mathcal{D}_A(x,\qt)\Theta(P_\perp^2-\qt^2)\\
    &= \frac{\alpha_s}{3\pi} \sum_{i=q} e_i^2 \times \frac{\alpha_s}{2\pi^2}\frac{1}{P_\perp^2}\int_x^1\der\xi P_{gg}(\xi)\frac{x}{\xi}G\left(\frac{x}{\xi},P_\perp^2\right) \,, \label{eq:FUUL-finite-dilute}
\end{align}
with
\begin{align}
    P_{gg}(\xi)=\frac{2N_c[1-\xi(1-\xi)]^2}{\xi(1-\xi)_+}\,,\quad xG(x,P_\perp^2)=\frac{C_F}{2\pi^2\alpha_s }\int_{\Bt,\qt}\qt^2\mathcal{D}_A(x,\qt)\Theta(P_\perp^2-\qt^2) \,.
\end{align}
In the expression for $P_{gg}$, the plus prescription is defined by
\begin{align}
    \int_x^1\der \xi \  \frac{f(\xi)}{(1-\xi)_+}&=\int_x^1\der\xi \left[\frac{f(\xi)-f(1)}{1-\xi}\right]\,,
\end{align}
for any function $f(\xi)$ with a well defined limit as $\xi\to 1$.
In the second line of Eq.\,\eqref{eq:FUUL-finite-dilute}, we have recognized in the numerical factor $2N_c\ln(1/z_0)-11 N_c/3$ the integral of the $g\to gg$ splitting function $P_{gg}(\xi)$ when $z_0=x\Pt^2/Q^2$. We have also set the longitudinal scale of the gluon PDF to $x/\xi$ to make manifest that the dilute limit corresponds to a single $g\to gg$ DGLAP step with an incoming gluon from the target wave-function. Strictly speaking, the exact $x$ value of the gluon PDF is not under controlled in our CGC calculation; its determination requires including some sub-eikonal corrections~\cite{Altinoluk:2020oyd,Altinoluk:2021lvu,Altinoluk:2022jkk} as discussed in \cite{Boussarie:2020fpb,Boussarie:2021wkn,Boussarie:2023xun} in the context of inclusive DIS and the exclusive Compton process and \cite{Fu:2023jqv,Fu:2024sba} in the context of the matching between the CGC effective theory and the high-twist formalism.

Finally, the overall factor $\alpha_s/(3\pi)$ in Eq.\,\eqref{eq:FUUL-finite-dilute} is the hard factor for the $\gamma^*_L+g$ channel in agreement with the Altarelli-Martinelli identity discussed in section A of this Supplemental Material.

\end{widetext}
\end{document}